\documentstyle[aps,epsf,graphicx]{revtex}
\def\xvec{{\bf x}}

\def\xp{x^\prime}
\def\xpvec{\xvec^\prime}
\def\Fvec{{\bf F}}
\def\uvec{{\bf u}}
\def\nuvec{{\mbox{\boldmath $\nu$}}}
\def\omvec{{\mbox{\boldmath $\omega$}}}
\def\Omvec{{\mbox{\boldmath $\Omega$}}}
\def\crossprod{{\mbox{\boldmath $\times$}}}
\def\dotprod{{\mbox{\boldmath $\cdot$}}}
\def\xivec{{\mbox{\boldmath $\xi$}}}
\def\grad{{\mbox{\boldmath $\nabla$}}}
\def\curl{\grad\crossprod}

\def\be{\begin{equation}}
\def\ee{\end{equation}}
\def\baray{\begin{eqnarray}}
\def\earay{\end{eqnarray}}
\def\bp{\beta^\prime}
\def\bdp{\beta^{\prime\prime}}

\begin{document}

\def\bxi{{\mbox{\boldmath $\xi$}}}
\def\bnabla{{\mbox{\boldmath $\nabla$}}}
\def\bomega{{\mbox{\boldmath $\omega$}}}
\def\bOmega{{\mbox{\boldmath $\Omega$}}}
\def\bnu{{\mbox{\boldmath $\nu$}}}
\newcommand{\hrho}{{\hat \varrho}}
\newcommand{\hf}{\hat{\bf f}}
\newcommand{\vv}{{\bf v}}
\newcommand{\hv}{{\hat {\bf v}}}
\newcommand{\hu}{\hat {\bf u}}
\newcommand{\bu}{{ u}}
\newcommand{\tr}{{\rm ~Tr}}
\newcommand{\hV}{{\hat V}}
\newcommand{\Comma}{{,}}
\newcommand{\Point}{{.}}

\title{Perturbations of self-gravitating, ellipsoidal superfluid-normal 
fluid mixtures}

\author{A. Sedrakian}
\address
{Kernfysisch Versneller Instituut,
NL-9747 AA Groningen, The Netherlands,\thanks{Postal address.} \\
and Institute for Nuclear Theory,
University of Washington, Seattle WA 98195-1550
}
\author{I. Wasserman}
\address
{Center for Radiophysics and Space Research,
                Cornell University, Ithaca, NY 14853}

\maketitle

\begin{abstract}
We study the perturbation modes of rotating superfluid ellipsoidal
figures of equilibrium in the framework of the two-fluid superfluid
hydrodynamics and Newtonian gravity. Our calculations focus on
linear perturbations of background equilibria in which the two fluids
move together, the total density is uniform, and the densities of the
two components are proportional to one another, with ratios that are
independent of position. The motions of the two fluids are coupled by
mutual friction, as formulated by Khalatnikov. We show that there are
two general classes of modes for small perturbations: one class
in which the two fluids move together and the other in which there
is relative motion between them. The former are identical to the
modes found for a single fluid, except that the rate of viscous
dissipation, when computed in the secular (or ``low Reynolds number'')
approximation under the assumption of a constant kinematic viscosity, 
is diminished by a factor $f_N$, the fraction of the
total mass in the normal fluid. The relative modes are completely new,
and are studied in detail for a  range of values for the
phenomenological mutual friction coefficients, relative densities of
the superfluid and normal components, and, for Roche ellipsoids, binary
mass ratios. We find that there are no new secular
instabilities connected with the relative motions of the two fluid
components. Moreover, although the new modes are subject to viscous
dissipation (a consequence of viscosity of the normal matter), they do
not emit gravitational radiation at all.
\end{abstract}

\section{Introduction}

The problem of the equilibrium and stability of rotating neutron stars
is encountered in various astrophysical contexts, ranging from the
limiting frequencies of rapidly rotating isolated millisecond pulsars,
emission of gravitational waves in neutron star-neutron star
or neutron star-black hole binaries, to the generation of  $\gamma$-rays
in the neutron star mergers and $X$-rays in
accreting systems\cite{CONF}. Considerable current
interest is attached to the
problem of neutron star binary inspiral, which would be the primary
source of gravitational wave radiation for detection by future
laser interferometers. Such a detection, apart from testing
the general theory of relativity,  potentially could
provide useful information on the equation of
state of superdense matter. Also the stability criteria for rapidly
rotating neutron stars are essential for placing firm upper limits
on the frequencies to which millisecond pulsars can be spun up
thereby constraining the range of admissible equations of states.

High precision, fully relativistic treatments of rapidly rotating
isolated neutron stars and binaries comprising two neutron stars
have become available in recent years\cite{CONF}.
Nevertheless, the development of simpler models that provide a fast and
transparent insight into the underlying physics is needed
when  the basic set of equations is modified to include
new effects.

A systematic framework for the treatment of the equilibrium
and stability of rotating liquid masses bound by self-gravitation
in the Newtonian theory is contained in Chandrasekhar's
{\it Ellipsoidal Figures of Equilibrium}\cite{CHANDRA} (hereafter EFE).
The tensor virial method, developed most extensively by Chandrasekhar
and co-workers, transforms the local hydrodynamical equations
into global virial equations that contain the full
information on the structure and stability of the
Newtonian self-gravitating system as a whole.
The method describes, in a coherent manner,
the properties of solitary ellipsoids with and without intrinsic spin,
and ellipsoids in binaries subject to Newtonian tidal fields.
It is especially useful for studying divergence-free
displacements of uniform ellipsoids from equilibrium, in which
case the each perturbed virial equation yields (in the
absence of viscous dissipation) a different set of normal modes.

Recent alternative formulations of the theory of ellipsoids  are
based mainly on either the energy variation method \cite{ZELDOVICH,ST}
or the affine star model\cite{CARTER} or the (Eulerian) two potential
formalism\cite{IPSER}.
A large class of incompressible and compressible  ellipsoidal
models has been studied using these methods\cite{CARTER,IPSER,LRS}.
The energy variation method employs the observation that an
equilibrium configuration
is possible if the energy of the ellipsoid is an
extremum for variations of the ellipsoidal semiaxis at
a constant volume; the ellipsoidal figure is
stable only if the energy is a true local minimum.
In the affine star method the figures  are
described by a time-dependent Lagrangian as a  function of a deformation
matrix and its derivative. The structure of the star at any
particular time is related conformally to the initial unperturbed
sphere via a quadratic form constructed from the deformation matrix.

The purpose of this paper is to extend previous studies to a
treatment of the oscillation modes of ellipsoidal
figures of equilibrium that contain a mixture of normal fluid
and superfluid.
Many-body studies of the pair correlations in neutron star matter
show that the baryon fluids in their ground state form superfluid
condensates in the bulk of the star. The superfluid phases,
in the hydrodynamic limit, can be
treated as a mixture of superfluid condensate and normal matter.
The superfluid rotation is supported by the Feynman-Onsager vortex
lattice state, and on the average leads to quasi-rigid body rotation
of the superfluid component. The corresponding time-dependent
two-fluid hydrodynamic equations are completely
specified by two independent velocities for the superfluid and
the normal component and corresponding densities of the constituents.
The two fluids are coupled to one another by mutual friction
forces, which we model phenomenologically according the prescription
given by Khalatnikov \cite{KHALATNIKOV}. Because of the additional degrees
of freedom in this system, there are twice as many modes as for
a single fluid. A natural question is whether the new modes
affect the stability criteria previously deduced from studying
perturbations of a single self-gravitating fluid.

In our treatment of perturbations of a mixture of normal fluid
and superfluid, we shall follow most closely Chandrasekhar's
formulation. However, since the basic equations of motion for the
two fluids will include mutual friction between them, the
system we study is inherently dissipative. Nevertheless, since the
frictional forces only depend linearly on the relative velocity between
the two fluids and vanish in the background, where the two
fluids move together, one can still derive relations that
resemble Chandrasekhar's tensor virial equations. Because these
equations include dissipation, we shall prefer to regard them
as moments of the fluid equations, rather than tensor virial
equations. In fact, we shall relegate the derivation of perturbation
equations from these moment equations to the Appendix of the paper,
and instead derive the necessary equations for the fluid displacements
directly by taking moments of linearized equations of motion for
the two fluid components. In this paper, we concentrate on
two-fluid variants of the classical Maclaurin, Jacobi and
Roche ellipsoids.

Although our main aim is to access the oscillation modes and instabilities
of neutron stars within the ellipsoidal approximation,
the results obtained here may be of significance in other contexts
(Ref. \cite{CHANDRA}, the epilogue).
One example is the understanding of rapidly rotating nuclei
in the spirit of the Bohr-Wheeler model of a charged
incompressible liquid droplet\cite{BOHR}. In this case, the stability
is determined by the competition between the attractive surface
tension, the repulsive  Coulomb potential,  and the centrifugal stretching
due to the rotation\cite{COHEN,ROSEN}. Another example is the stability
of rotating superfluid liquid drops of Bose condensed atomic
gases, where the stability is determined through  an interplay among 
the pressure of the condensate, the confining potential of the
magnetic trap and the centrifugal potential~\cite{FETTER}.

Previous work on the oscillations of superfluid neutron
stars concentrated mainly on perturbations of  
non-rotating or slowly rotating isolated neutron stars
\cite{EPSTEIN,LINDBLOM,UMIN,COMER}, and used methods 
entirely different from the one adopted here. The propagation 
of acoustic waves in neutron star interiors, including those 
related to the relative motion of neutron-proton superfluids, 
was studied by Epstein\cite{EPSTEIN}, who found the compressional 
and shear modes related to short-wavelength oscillations of 
neutron star matter. 
The small-amplitude pulsation modes of superfluid neutron stars
were derived by Lindblom and Mendell \cite{LINDBLOM}, who
found that the lowest frequency modes were almost indistinguishable 
from the normal modes of a single fluid star. Their analytical 
solutions also reveal the existence of a spectrum of modes which 
are absent in a single fluid star. Subsequent work concentrated on 
numerical solutions for the radial and non-radial pulsations of 
the two-fluid stars  and identified distinct superfluid modes in 
the absence of rotation\cite{UMIN}. 
The r modes of slowly rotating two-fluid neutron stars
have  been derived  by Lindblom 
and Mendell \cite{LINDBLOM}, who find that they are identical to 
their ordinary-fluid counterparts to the lowest order in their
small-angular-velocity expansion. 
The linear oscillations of general relativistic stars composed of two 
non-interacting fluids in a non-rotating static background have been 
studied by Comer {\it et al.} \cite{COMER}.

Our calculations allow arbitrary fast rotation, in the context of 
(incompressible) Newtonian fluid models. 
One may anticipate that the effects of superfluidity  
on oscillation modes,  if any, should be affected by the underlying 
vortex structure of the rotating superfluid. In our treatment, 
dissipation arises because of the drag forces experienced by the
vortex lines as they move through the normal fluid (and there 
is no dissipation if the drag force is zero).
We ignore the motions related to the isospin degrees of freedom 
in the core of a neutron star and, hence, the mutual entrainment of the 
neutron and proton condensates, as well as forces arising from deviations from 
$\beta$ equilibrium.  The two-fluid equations used in the remainder of 
this work can adequately describe the mutual friction of a two-condensate 
fluid in the core of a neutron star, since the entrainment effect 
renormalizes the effective superfluid densities and the frictional
coefficients, i.e. the phenomenological input in the two-fluid equations.
While the ellipsoidal  approximation to a superfluid neutron star
is restrictive, it allows us to
study the effects of vorticity on the oscillation modes of
a self-gravitating star in a transparent manner, avoiding 
complications due to the star's inhomogeneity (multi-layer 
composition)\cite{RUDERMAN}.

\section{Perturbation Equations}
\label{ve2pe}

The equations of motion for a mixture of two fluids may be
summarized simply as
\be
\rho_\alpha D_\alpha u_{\alpha,i}=-{\partial p_\alpha\over
\partial x_i}-\rho_\alpha{\partial\phi\over\partial x_i}
+{1\over 2}\rho_\alpha{\partial\vert\Omvec\crossprod\xvec\vert^2
\over\partial x_i}+2\rho_\alpha\epsilon_{ilm}u_{\alpha,l}\Omega_m
+F_{\alpha\beta,i},
\label{eq:euler}
\ee
where the subscript $\alpha\in\{S,N\}$ identifies the fluid component,
and Latin subscripts denote coordinate directions; $\rho_\alpha$,
$p_\alpha$, and ${\bf u}_\alpha$ are the density, pressure, and
velocity of fluid $\alpha$, $\phi$ is the gravitational potential, and
${\bf F}_{\alpha\beta}$ is the {\it mutual friction force} on fluid
$\alpha$ due to fluid $\beta$. These equations
have been written in a frame rotating with angular velocity
$\Omvec$ relative to some inertial coordinate reference system.
The total time derivative operator
\be
D_\alpha\equiv{\partial\over\partial t}+u_{\alpha,j}
{\partial\over\partial x_j}.
\ee
The gravitational potential $\phi$ is derived from
\be
\nabla^2\phi=\nabla^2(\phi_S+\phi_N)=4\pi G[\rho_S(\xvec)+\rho_N(\xvec)];
\ee
the individual fluid potentials $\phi_\alpha$ obey
$\nabla^2\phi_\alpha=4\pi G\rho_\alpha$.
The two fluids are coupled to one another via the frictional force
$\Fvec_{\alpha\beta}$ which is antisymmetric on interchange of
$\alpha$ and $\beta$. For a normal-superfluid mixture
\be
\Fvec_{SN}=-\Fvec_{NS}\equiv
\rho_S\omega_S\{\bp\nu\crossprod(\uvec_S-\uvec_N)
+\beta\nuvec\crossprod[\nuvec\crossprod(\uvec_S-\uvec_N)]
-\bdp\nuvec\dotprod(\uvec_S-\uvec_N)\},
\label{eq:fdef}
\ee
where $\beta$, $\bp$ and $\bdp$ are coupling coefficients, and
$\omvec_S=\nuvec\omega_S\equiv\curl\uvec_S$; in components
we have
\be
F_{SN,i}=-\rho_S\omega_S\beta_{ij}(u_{S,j}-u_{N,j}),
\ee
where, from Eq. (\ref{eq:fdef}),
\be
\beta_{ij}=\beta\delta_{ij}+\beta^\prime\epsilon_{ijm}\nu_m
+(\bdp-\beta)\nu_i\nu_j.
\label{eq:betaijdef}
\ee
The net rate at
which this force does work is
\be
\uvec_S\dotprod\Fvec_{SN}+\uvec_N\dotprod\Fvec_{NS}
=-\rho_S\omega_S\{\beta\vert\nuvec\crossprod(\uvec_S-\uvec_N)\vert^2
-\bdp[\nuvec\dotprod(\uvec_S-\uvec_N)]^2\};
\ee
there is no dissipation associated with the term proportional
to $\bp$ in $\Fvec_{\alpha\beta}$. Throughout this paper, we assume
that $\beta$, $\beta^\prime$ and $\beta^{\prime\prime}$ are independent
of position in the fluid mixture. 
In Eq. (\ref{eq:fdef}), we have neglected the effects of the vortex
tension, and expressed the mutual friction force in terms of the
phenomenological coefficients $\beta$, $\beta^\prime$ and
$\beta^{\prime\prime}$. While these parameters determine the
macroscopic behavior of the fluid system, they are not the optimal
ones for connecting microscopic parameters of the mixture to
its macroscopic motion. Instead, the macroscopic results can
be parametrized in terms of frictional coefficients $\eta$
and $\eta^\prime$, which connect $\beta$ and $\beta^\prime$
to the drag on individual superfluid vortices via the relations
\be
\beta={\eta\rho_S\omega_S\over\eta^2+(\rho_S\omega_S-\eta^\prime)^2}
\qquad
\beta^\prime=1-{\rho_S\omega_S(\rho_S\omega_S-\eta^\prime)\over
\eta^2+(\rho_S\omega_S-\eta^\prime)^2}.
\label{BETA_ETA}
\ee
The physical meaning of $\eta$'s is apparent from the equation of 
motion of a single vortex line
\be\label{1.4.2}
\rho_S\omega_S\left[\left(\uvec_S-\uvec_L\right)
\times{\nuvec}\right] - \eta\left(\uvec_L-\uvec_N\right)
+\eta'\left[\left(\uvec_L-\uvec_N\right)\times{\nuvec}\right] = 0,
\label{FORCE_BALANCE}
\ee
where ${\uvec}_L$ is the vortex velocity.
Equation (\ref{FORCE_BALANCE}) states that the Magnus force, which represents 
a lifting force due to the superflow imposed on the vortex circulation,
is balanced by the {\it viscous friction forces} 
along the vortex motion (the term
$\propto \eta$)  and perpendicular to the vortex motion (the term
$\propto \eta'$); these latter forces
arise from the scattering of the normal quasiparticles
off the vortex line\footnote{The inertial mass of the vortex is neglected
in the standard formulation of the two-fluid superfluid hydrodynamics
\cite{KHALATNIKOV}.}. The characteristic dynamical relaxation 
time scale related to the vortex motion can be defined as 
\be
\tau_{\rm D} = \frac{1}{\langle\omega_S\rangle} \left(\frac{\eta}{\rho_S\omega_S}
+\frac{\rho_S\omega_S}{\eta} \right),
\label{TAU_D}
\ee  
where $\langle \omega_S \rangle$ is the superfluid circulation
averaged over macroscopic scales; e.g. for uniformly rotating 
superfluid $\langle \omega_S \rangle = 2\Omega_S$, where 
$\Omega_S$ is the superfluid rotation frequency.
For fixed density $\rho_S$,
$\tau_D\to\infty$  asymptotically,   when
$\eta\gg\rho_S\omega_S$ (strong coupling limit) 
{\it and} $\rho_S\omega_S\ll\eta$ (weak coupling limit). Its  
minimal value is attained when $\eta/\rho_S\omega_S = 1$. 

The relations (\ref{BETA_ETA}) 
do not determine $\beta^{\prime\prime}$, which, if
nonzero, implies friction {\it along} vortex lines, which would be
possible if vortices oscillate or are deformed in the plane perpendicular
to the rotation axis. Generally, we assume in this paper that
$\beta^{\prime\prime}\ll\beta$ and $\beta^\prime$, but occasionally
we retain nonzero (and not necessarily negligible) $\beta^{\prime\prime}$
to examine its effects on the modes.
Moreover, if we assume that the drag force on a vortex line is
principally opposite to its velocity relative to the normal fluid,
then $\eta^\prime\ll\eta$, and the relationships between $\beta$
and $\beta^\prime$ and the drag force on vortex lines simplifies
accordingly.

Ultimately, we want to compute perturbations of Eq. (\ref{eq:euler})
around some presumed background state. In general, the two fluids
need not occupy the same volume, and we shall suppose that
fluid $\alpha$ occupies a volume $V_\alpha$. However, we shall
restrict ourselves to background states in which the two
fluids occupy identical volumes and have densities $\rho_\alpha(\xvec)
=f_\alpha\rho(\xvec)$, where $\rho(\xvec)$ is the total density
and $f_\alpha$ does not depend on $\xvec$. The perturbation equations
derived below can be applied to nonuniform background states satisfying
these conditions, but we shall only consider uniformly dense backgrounds,
as was done by Chandrasekhar. Note, though, that we shall not assume
that the two fluids must occupy the same volumes in the perturbed
state. In fact,  at least for low order perturbations,
there are no nontrivial perturbations that leave the volumes of the
two fluids identical.

Following the example set by Chandrasekhar, we could take moments of
the fluid equations (\ref{eq:euler}) to obtain tensor virial
theorems of various orders, and then perturb them to find
linear modes for uniform ellipsoids. We have derived the
necessary moment equations in this way (see the Appendix),
by analogy to Chandrasekhar's treatment
for a single fluid, but here we present a somewhat different
(and possibly more transparent approach) to their derivation.
We begin with the equation of motion for the
displacement of fluid $\alpha$ from equilibrium, a direct generalization
of Eq. (107) in Chap. 2, Sec. 14 in EFE:
\baray
\rho_\alpha{d^2\xi_{\alpha,i}\over dt^2}&=&
-{\partial\xi_{\alpha,l}\over\partial x_l}{\partial p_\alpha\over\partial x_i}
-{\partial\Delta_\alpha p_\alpha\over\partial x_i}-\rho_\alpha{\partial\Delta_\alpha\phi
\over\partial x_i}
+\rho_\alpha{\partial\over\partial x_i}\biggl[{\xi_{\alpha,l}\over 2}
\biggl({\partial\vert\Omvec\crossprod\xvec\vert^2
\over\partial x_l}\biggr)\biggr]+2\rho_\alpha\epsilon_{ilm}\Omega_m{d\xi_{\alpha,l}\over dt}
\nonumber\\& &
+{\partial\xi_{\alpha,l}\over\partial x_i}\biggl({\partial p_\alpha\over\partial x_l}
+\rho_\alpha{\partial\phi\over\partial x_l}-{\rho_\alpha\over 2}
{\partial\vert\Omvec\crossprod\xvec\vert^2\over\partial x_l}\biggr)
+F_{\alpha\beta,i},
\label{alphapert}
\earay
where the Lagrangian variation $\Delta_\alpha Q$ denotes the change in $Q$ seen by
a moving element of fluid $\alpha$; in particular, it is easy to show that
$\Delta_\alpha\phi(\xvec)$ separates
into an Eulerian part, $\delta\phi$, plus $\xi_{\alpha,l}{\partial\phi/
\partial x_l}$, where
\be
\delta\phi\equiv -G\sum_{\gamma=\alpha,\beta}\int_V{d^3\xp
\rho_\gamma(\xpvec)\xi_{\gamma,l}(\xpvec)
{\partial\over\partial\xp_l}\biggl({1\over
\vert\xvec-\xpvec\vert}\biggr)}
\label{eq:eulerphi}
\ee
is the Eulerian potential perturbation.
We assume that the mutual friction force is
$F_{\alpha\beta,i}=0$ in the background solution, i.e., either the two fluids
are stationary in the rotating frame for the background or have identical
fluid velocities in this frame. Our strategy will be to take moments of Eq.
(\ref{alphapert}) by multiplying by appropriate factors of $x_i$ and integrating
over the (common) volume of the unperturbed background configuration.
Note that this does not restrict the volumes of the perturbed fluids in any way.

Although the method we shall use to derive the perturbations resembles
Chandrasekhar's, there is an important distinction due to the mutual friction
force. Chandrasekhar's method of solution yields exact modes only in the
dissipationless limit, where the modes of uniform ellipsoids are, respectively,
linear, quadratic, cubic, etc. functions of the coordinates. Viscous terms
would prevent exact solution in this manner, and one resorts to an approximation
in which they are evaluated by substitution of the inviscid eigenfunctions
(EFE, Chap. 5, \S37(b)). However, it is possible to employ the moment
method to find exact modes for the normal fluid-superfluid mixture
coupled by mutual friction because
${\bf F}_{\alpha\beta}$ is a linear function of the velocity difference
between the two fluids, not of their spatial derivatives (as is the case
for viscous dissipation). By taking moments, we can derive the analogue
of tensor virial theorems of various orders, but because the equations
involve manifestly dissipative mutual friction forces, we prefer to think
of these merely as moments of the original equations of motion. Of course,
if we also choose to include the effects of viscous dissipation in the
normal fluid, which we shall not do in this paper, we must resort to
a low Reynolds number approximation, as was done in EFE. We discuss this
briefly in Sec. \ref{viscous}.

Using the above result for the perturbed gravitational potential, we can
rewrite Eq. (\ref{alphapert}) as
\baray
\rho_\alpha{d^2\xi_{\alpha,i}\over dt^2}&=&
-{\partial\xi_{\alpha,l}\over\partial x_l}{\partial p_\alpha\over\partial x_i}
-\xi_{\alpha,l}{\partial^2p_\alpha\over\partial x_i\partial x_l}
-{\partial\delta p_\alpha\over\partial x_i}-\rho_\alpha{\partial\delta\phi\over
\partial x_i}
\nonumber\\
& & -\rho_\alpha\xi_{\alpha,l}{\partial^2\phi\over\partial x_i\partial x_l}
+\rho_\alpha\xi_{\alpha,l}(\Omega^2\delta_{il}-\Omega_i\Omega_l)
+2\rho_\alpha\epsilon_{ilm}\Omega_m{d\xi_{\alpha,l}\over dt}
+F_{\alpha\beta,i}.
\earay
Then if we define
\be
\xivec_+=f_S\xivec_S+f_N\xivec_N\qquad
\xivec_-=\xivec_S-\xivec_N,
\ee
we find
\baray
\rho{d^2\xi_{+,i}\over dt^2}&=&
-{\partial\xi_{+,l}\over\partial x_l}{\partial p\over\partial x_i}
-\xi_{+,l}{\partial^2p\over\partial x_i\partial x_l}
-{\partial\delta p\over\partial x_i}-\rho{\partial\delta\phi\over
\partial x_i}
\nonumber\\
& & -\rho\xi_{+,l}{\partial^2\phi\over\partial x_i\partial x_l}
+\rho\xi_{+,l}(\Omega^2\delta_{il}-\Omega_i\Omega_l)
+2\rho\epsilon_{ilm}\Omega_m{d\xi_{+,l}\over dt}
\label{eq:common}
\earay
where $\delta p=\delta p_S+\delta p_N$, and
\baray
\rho{d^2\xi_{-,i}\over dt^2}&=&
-{\partial\xi_{-,l}\over\partial x_l}{\partial p\over\partial x_i}
-\xi_{-,l}{\partial^2p\over\partial x_i\partial x_l}
-{1\over f_S}{\partial\delta p_S\over\partial x_i}
+{1\over f_N}{\partial\delta p_N\over\partial x_i}
\nonumber\\
& & +\rho\xi_{-,l}(\Omega^2\delta_{il}-\Omega_i\Omega_l)
-\rho\xi_{-,l}{\partial^2\phi\over\partial x_i\partial x_l}
+2\rho\epsilon_{ilm}\Omega_m{d\xi_{-,l}\over dt}
\nonumber\\& &
-\rho\omega_S\biggl(1+{f_S\over f_N}\biggr)\beta_{ik}{d\xi_{-,k}\over dt}.
\label{eq:different}
\earay
Equation (\ref{eq:common}) is identical to what is found for a single fluid,
and therefore contains the well-known modes documented by Chandrasekhar:
if we define
\be
V_{i;j}=\int_V{d^3x\rho\xi_{+,i}x_j},
\ee
then we find
\baray
{d^2V_{i;j}\over dt^2}&=&2\epsilon_{ilm}\Omega_m{dV_{l;j}\over dt}
+\Omega^2V_{ij}-\Omega_i\Omega_kV_{kj}+\delta_{ij}\delta\Pi
\nonumber\\
&-&\pi G\rho\biggl(2B_{ij}V_{ij}-a_i^2\delta_{ij}\sum_{l=1}^3A_{il}V_{ll}
\biggr),
\label{meanflow_basic}
\earay
where $\delta\Pi\equiv\delta\Pi_S+\delta\Pi_N$ and all other quantities
are defined exactly as in EFE.
All of the new modes of a mixture of normal fluid and superfluid are
contained in Eq. (\ref{eq:different}) for their relative displacements.
One noteworthy feature of
Eq. (\ref{eq:different}) is that the Eulerian gravitational potential
does not appear. Consequently, the new normal modes of the system
only depend on the unperturbed gravitational potential; for perturbations
of homogeneous ellipsoids,
only the coefficients $A_i$ defined by EFE, Chap. 3, Eqs. (18)
and (40), will appear.

For the most part, we shall be interested in displacements that are
linear functions of $x_i$ in this paper. For the homogeneous ellipsoids,
we can find the new modes that result from the differential displacements
of normal fluid and superfluid by taking the first moment of Eq.
(\ref{eq:different}), e.g. by multiplying by $x_j$ and integrating over
the unperturbed volume.\footnote{In the nomenclature of EFE, this is
the second order virial equation corresponding to Eq. (\ref{eq:different}).}
If we define
\be
U_{i;j}=\int_V{d^3x\rho\xi_{-,i}x_j},
\ee
then we find
\baray
{d^2U_{i;j}\over dt^2}&=&
2\epsilon_{ilm}\Omega_m{dU_{l;j}\over dt}
+\Omega^2U_{ij}-\Omega_i\Omega_kU_{kj}+\delta_{ij}\biggl({\delta\Pi_S\over f_S}
-{\delta\Pi_N\over f_N}\biggr)
\nonumber\\
&-&2\pi G\rho A_iU_{ij}-\omega_S\biggl(1+{f_S\over f_N}\biggr)
\beta_{ik}{dU_{k;j}\over dt},
\label{eq:differentmom1}
\earay
where
\be
\delta\Pi_\alpha\equiv\int_V{d^3x\delta p_\alpha}.
\ee
To obtain Eq. (\ref{eq:differentmom1}), various surface terms can be eliminated
using the conditions that $p_\alpha$ and $\Delta_\alpha p_\alpha=\delta p_\alpha
+\xi_{\alpha,l}\partial p_\alpha/\partial x_l$ vanish on the boundary; also,
the equation of hydrostatic equilibrium for the unperturbed configuration,
\be
0={\partial p\over\partial x_i}+\rho{\partial\phi\over\partial x_i}
-\rho{\partial\over\partial x_i}\biggl({\vert\Omvec\crossprod\xvec\vert^2
\over 2}\biggr),
\ee
must be used. It is straightforward to compute higher moments of Eq.
(\ref{eq:different}). For example, taking its second 
moment\footnote{This is the third order virial equation 
in the nomenclature of EFE.}
by multiplying by $x_jx_k$ and integrating over the unperturbed 
volume gives
\baray
{d^2U_{i;jk}\over dt^2}&=&\delta_{ij}\biggl({\delta\Pi_{S,k}\over f_S}
-{\delta\Pi_{N,k}\over f_N}\biggr)+\delta_{ik}
\biggl({\delta\Pi_{S,j}\over f_S}
-{\delta\Pi_{N,j}\over f_N}\biggr)
\nonumber\\
& &-2\pi G\rho A_iU_{ijk}
+(\Omega^2\delta_{il}-\Omega_i\Omega_l)U_{ljk}
\nonumber\\& &
+\biggl[2\epsilon_{ilm}\Omega_m-\omega_S\biggl(1+{f_S\over f_N}\biggr)
\beta_{il}\biggr]{dU_{l;jk}\over dt},
\label{eq:differentmom2}
\earay
where, by analogy to definitions in EFE for a single fluid,
\baray
U_{i;jk}&=&\int_V{d^3x\rho\xi_{-,i}x_jx_k}\nonumber\\
U_{ijk}&=&U_{i;jk}+U_{j;ki}+U_{k;ij}\nonumber\\
\delta\Pi_{\alpha,k}&=&\int_V{d^3x x_k\delta p_\alpha}.
\earay
Equations (\ref{eq:differentmom1}) and (\ref{eq:differentmom2})
only apply to unperturbed states
that are static in the rotating frame.

\section{New Modes of a Two-Fluid Mixture}

In this section, we derive the characteristic equations for the normal
modes for relative fluid displacements implied by Eq. (\ref{eq:differentmom1})
assuming $U_{i;j}\propto\exp(\lambda t)$. We list the results separately for
perturbations of Maclaurin, Jacobi and Roche ellipsoids.
Note that the modes described by  Eq. (\ref{meanflow_basic}) 
are identical to those treated in EFE.

The superfluid state of the matter does not
affect the equilibrium figure.\footnote{Here, we 
have taken $p_\alpha=f_\alpha p$ in the background
state, which is a mathematically convenient idealization. More realistically,
contributions from the pair condensation energy and the energy density of
the superfluid vortex lattice, which distinguish the superfluid from the
normal fluid, could play a role in both the equilibria and the perturbations.
We need not restrict our attention to displacements for which the pressure
perturbations of the two fluids 
are still proportional to one another, although,
as argued in \S\ref{viscous}, for the solenoidal displacements considered
here, this turns out to be the case.} In equilibrium,
Eq. (\ref{eq:differentmom1}) is satisfied trivially, for all virials
$U_{ij} = 0$ in the absence of relative motion between the superfluid
and normal components. The equilibrium figure follows from
Eq. (\ref{meanflow_basic}) by dropping the temporal variations and is
identical to its non-superfluid counterpart.

For  {\it irrotational} ellipsoids
we can set $\omega_S = 2 \Omega$ in Eq. (\ref{eq:differentmom1}) and
using the compact notations
\be
\delta\tilde\Pi =  \biggl({\delta\Pi_S\over f_S}
-{\delta\Pi_N\over f_N}\biggr),
\quad \tilde\beta_{ik} =\biggl(1+{f_S\over f_N}\biggr)\,\beta_{ik}\Comma
\ee
we rewrite the Eq. (\ref{eq:diffflow}) as
\baray
{d^2U_{i;j}\over dt^2}&=&2\epsilon_{ilm}\Omega_m{dU_{l;j}\over dt}
+\Omega^2U_{ij}-\Omega_i\Omega_kU_{kj}+\delta_{ij} \delta\tilde\Pi
-2\pi G\rho A_iU_{ij}-2\Omega
\tilde\beta_{ik}{dU_{k;j}\over dt}\Point
\label{eq:diffflow}
\earay
For time-dependent Lagrangian displacements of the form
\be
{\xivec}_{\alpha}( x_i, t) = {\xivec}_{\alpha}( x_i)e^{\lambda t}\Comma
\ee
Eq. (\ref{eq:diffflow}) becomes
\baray
\lambda^2 U_{i;j}-2\epsilon_{ilm}\Omega_m \lambda U_{l;j} &=&
+\Omega^2U_{ij}-\Omega_i\Omega_kU_{kj}+(\pi\rho G)^{-1}\delta_{ij}
\delta\tilde\Pi-2 A_iU_{ij}-2\Omega\lambda
\tilde\beta_{ik}U_{k;j};
\label{eq:modes}
\earay
here all frequencies are measured in the units $(\pi \rho G)^{1/2}$.
Equation (\ref{eq:modes}) contains all the second harmonic
modes of the relative oscillation of Maclaurin and Jacobi ellipsoids, and
only requires a slight modification for the application to Roche ellipsoids.

\subsection{Superfluid Maclaurin spheroid}

Next, we specialize Eq. 
(\ref{eq:diffflow}) to
Maclaurin spheroids,
the  equilibrium figures of a self-gravitating fluid
with two equal semi-major axis, say $a_1$ and $a_2$,
uniformly rotating about the third semi-major axis $a_3$
(i.e. the $x_3$ axis).
The sequence of quasi-equilibrium figures
can be parametrized by the eccentricity
$\epsilon^2 =1-a_3^2/a_1^2$, with (squared) angular velocity
$\Omega^2 = 2 \epsilon^2 B_{13}$, in units of $(\pi\rho G)^{1/2}$.

Surface deformations related to various modes can be
classified by corresponding terms of the
expansion in surface harmonics labeled by the indexes $l,m$.
Second order harmonic deformations correspond to $l=2$
with five distinct values of $m$, $-2\le m \le 2$.
The 18 equations represented by
Eq. (\ref{eq:diffflow}) separate into two independent
subsets which  are odd and even with respect to index 3.
The corresponding oscillation modes can be treated separately.

\subsubsection{Relative transverse shear modes}
These modes correspond
to surface deformations with $\vert m\vert = 1$
and represent relative shearing of the northern
and southern hemispheres of the ellipsoid.
The components of Eq. (\ref{eq:modes}), which are odd in
index 3, are
\baray\label{mac_modes:1.1}
\lambda^2 U_{3;1} &=& -2 A_{3} U_{31}
-2 \Omega\tilde\beta''\lambda U_{3;1}\Comma\\
   \label{mac_modes:1.2}
\lambda^2 U_{3;2} &=& -2 A_{3} U_{32}
-2 \Omega\tilde\beta''\lambda U_{3;2}\Comma\\
   \label{mac_modes:1.3}
\lambda^2 U_{1;3} - 2 \Omega\lambda U_{2;3}
&=&\left(-2 A_{1}+\Omega^2\right)U_{13}
-2\Omega\tilde\beta\lambda U_{1;3}
- 2 \Omega\tilde\beta'\lambda U_{2;3}\Comma\\
   \label{mac_modes:1.4}
\lambda^2 U_{2;3} + 2 \Omega\lambda U_{1;3}
&=&\left(-2 A_{1}+\Omega^2\right)U_{23}
-2\Omega\tilde\beta\lambda U_{2;3}
+2 \Omega\tilde\beta'\lambda U_{1;3}\Point
\earay
Note that because of the degeneracy in indexes 1
and 2 for the Maclaurin spheroid
$A_1 = A_2$.
We sum  Eqs. (\ref{mac_modes:1.1}), (\ref{mac_modes:1.3})
and (\ref{mac_modes:1.2}), (\ref{mac_modes:1.4}), respectively, and
use the symmetry properties of $U_{ij}$ combined with Eqs.
(\ref{mac_modes:1.2}), (\ref{mac_modes:1.4}). We find
\baray
&&\left[\left(\lambda^2+2\Omega\tilde\beta''\lambda\right)
\left(\lambda^2+2\Omega\tilde\beta\lambda\right)+
2 \left(\lambda^2+2\Omega\tilde\beta\lambda\right) A_3
+ \left(\lambda^2+2\Omega\tilde\beta''\lambda\right) (2 A_1-\Omega^2)
\right] U_{13} \nonumber \\
&&\hspace{4cm} -
2\Omega\lambda(1-\tilde\beta')
\left(\lambda^2+2\Omega\tilde\beta''\lambda + 2A_3\right)
 U_{23} = 0\Comma
\label{mac_modes:1.5}\\
&&\left[\left(\lambda^2+2\Omega\tilde\beta''\lambda\right)
\left(\lambda^2+2\Omega\tilde\beta\lambda\right)+
2 \left(\lambda^2+2\Omega\tilde\beta\lambda\right) A_3
+ \left(\lambda^2+2\Omega\tilde\beta''\lambda\right) (2 A_1-\Omega^2)
\right] U_{23} \nonumber \\
&&\hspace{4cm} +
2\Omega\lambda(1-\tilde\beta')
\left(\lambda^2+2\Omega\tilde\beta''\lambda + 2A_3\right)
 U_{13} = 0\Point
\label{mac_modes:1.6}
\earay
It is instructive to consider first the limit of zero mutual friction, in which
case Eqs. (\ref{mac_modes:1.5})-(\ref{mac_modes:1.6}) reduce to
\baray\label{mac_modes_bis:1.5}
 &&\lambda\left[\lambda^2 + (2 A_1 + 2 A_3-\Omega^2)
\right] U_{13} - 2\Omega\left(\lambda^2+ 2A_3\right)
 U_{23} = 0\Comma \nonumber \\
     \label{mac_modes_bis:1.6}
&&\lambda \left[\lambda^2+  (2 A_1 + 2 A_3-\Omega^2)
\right] U_{23} + 2\Omega\left(\lambda^2 + 2A_3\right)
 U_{13} = 0,
\label{macmodesnondiss1}
\earay
excluding the trivial mode $\lambda = 0$.  The characteristic
equation can be factorized by substituting $\lambda = i\sigma$
\be
\sigma\left[\sigma^2 - 2(A_1+A_3) +\Omega^2\right]\pm
2\Omega (\sigma^2 - 2A_3) = 0\Point
\ee
The purely rotational mode $\sigma=\Omega$
decouples only in the spherical symmetric limit where $A_1 = A_3$.
If only axial symmetry is imposed then
the characteristic equation is
third order:
\baray
\sigma^3  \pm 2\Omega{\sigma^2} +
 \left[ -2\,\left(A_1 + A_3\right)  + \Omega^2 \right]\sigma
  \mp 4\,A_3\,\Omega=0.
\earay
Along the entire sequence parametrized in terms of the
eccentricity the three modes are real
\footnote{This is easy to prove directly from Eq. (\ref{macmodesnondiss1}).
Write the dispersion relation as $f(\lambda^2)=0$. Then show that
(i) $f(\lambda^2)\to\pm\infty$ as $\lambda^2\to\pm\infty$, (ii) $f(0)>0$,
and (iii) the two extrema of $f(\lambda^2)$ are both at $\lambda^2<0$. Thus,
the zeros of $f(\lambda^2)$ are all at $\lambda^2<0$, so $\sigma=i\lambda$
must be real.
}
and are given by
\baray\label{sig1}
\sigma_1 = \frac{2\,\Omega}{3} + (s_++s_-)\Comma \quad\quad
\sigma_{2,3} = \frac{2\,\Omega}{3} -\frac{1}{2} (s_+ +s_-)
    \pm \frac{i\sqrt{3}}{2} (s_+-s_-) \Comma
\earay
where
\baray
\label{sig_coff}
s_{\pm}^3 &=&
{\frac{8\,{{\Omega}^3}}{27}} +
  {\frac{2  \Omega  (A_1 - 2 A_3)-  \Omega^3 }{3}}\nonumber\\
&\pm& \left[{{{\left( -{\frac{4\,{{\Omega}^2}}{9}} -
           {\frac{2\,\left( A_1 + A_3 \right)
                   - {{\Omega}^2}}{3}} \right) }^3} +
      {{\left( {\frac{8\,{{\Omega}^3}}{27}} +
           {\frac{2 \Omega  (A_1 - 2 A_3)-
           \Omega^3 }{3}} \right) }^2}}\right]^{1/2}\Point
\earay
Three complementary modes follow from  Eqs.(\ref{sig1})-(\ref{sig_coff})
via the replacement  $\Omega\to -\Omega$. The frictionless modes
are real and are shown in Fig. 1. The two high frequency, frictionless
modes are roughly twice as large as the transverse shear modes of ordinary
Maclaurin spheroids. The third low frequency mode
corresponds to the nearly rotational mode and indeed coincides with $\Omega$
in the limits $\epsilon \to 0 $ and $A_1\to A_3$, but not generally.
In the dissipative case the characteristic equation is of
sixth order:
\baray \label{mac_modes:1.7}
&&\lambda^6 +  4 \Omega(\tilde\beta + \tilde\beta'') \lambda^5 +
    \Bigl[4 A_3 + 2\Omega(1-\tilde\beta')^2
   + 4\Omega^2 (\tilde\beta+\tilde\beta'')^2
  - 2(2A_1-\Omega^2)\Bigr]   \lambda^4
\nonumber\\
&&
   + \Bigl[16 A_3\tilde\beta\Omega + 8 A_3\tilde\beta''\Omega
   + 8\tilde\beta'' (1-\tilde\beta')^2\Omega^2
   + 16\tilde\beta^2\tilde\beta''\Omega^3  \nonumber\\
&&\hspace{3cm}  +
      16\tilde\beta\tilde\beta''^2\Omega^3
       - 4\tilde\beta\Omega (2A_1-\Omega^2)
       - 8\tilde\beta''\Omega (2A_1-\Omega^2)\Bigr]  \lambda^3
\nonumber\\
&&
   + \Bigl[4 A_3^2 + 8 A_3 \Omega(1-\tilde\beta')^2 + 16 A_3\tilde\beta^2\Omega^2
   + 32 A_3\tilde\beta\tilde\beta''\Omega^2 +
      8\tilde\beta''^2 (1-\tilde\beta')^2\Omega^3 + 16\tilde\beta^2\tilde\beta''^2\Omega^4
\nonumber\\
&&\hspace{3cm}
       - 4 A_3 (2A_1-\Omega^2) - 16\tilde\beta\tilde\beta''\Omega^2 (2A_1-\Omega^2) -
      8\tilde\beta''^2\Omega^2 (2A_1-\Omega^2)
      + (2A_1-\Omega^2)^2 \Bigr]    \lambda^2
\nonumber\\
&&
+\Bigl[16 A_3^2\tilde\beta\Omega + 16 A_3\tilde\beta'' (1-\tilde\beta')^2\Omega^2
+ 32 A_3\tilde\beta^2\tilde\beta''\Omega^3 - 8 A_3\tilde\beta\Omega (2A_1-\Omega^2)\nonumber\\
&&\hspace{3cm} -
      8 A_3\tilde\beta''\Omega (2A_1-\Omega^2) - 16\tilde\beta\tilde\beta''^2\Omega^3
      (2A_1-\Omega^2)
      + 4\tilde\beta''\Omega (2A_1-\Omega^2)^2 \Bigr]  \lambda
\nonumber\\
&&
      +\Bigl[8 A_3^2 \Omega(1-\tilde\beta')^2 + 16 A_3^2\tilde\beta^2\Omega^2
      - 16 A_3\tilde\beta\tilde\beta''\Omega^2 (2A_1-\Omega^2)
     + 4\tilde\beta''^2\Omega^2 (2A_1-\Omega^2)^2 \Bigr] = 0.
\earay
The real and imaginary parts of the relative transverse shear modes are shown
in the Fig. 1  for several values of $\eta$ and
$\eta^\prime=0=\beta^{\prime\prime}$ 
(here and below we scale $\eta$ in units of $\rho_S\omega_S$).
The real part of the low frequency rotational
modes is diminished as $\eta$ is increased;
the high frequency modes are unaffect except
in the stong coupling limit $\eta \ge 50$ [the strong and 
weak coupling limits are discussed after Eq. (\ref{TAU_D})].
The damping of the modes is maximal for $\eta = 1$ and 
decreases to zero for $\eta\to 0$ and $\eta\to\infty$. 
Note that in the limiting cases the vortices are locked either in the
superfluid ($\eta\to 0$) or the normal component ($\eta\to \infty$) and hence
the damping is ineffective. The transverse shear modes are stable for
arbitrary values of the eccentricity of the spheroid.

\subsubsection{Relative toroidal modes}
These modes correspond to $\vert m\vert = 2$ and
the motions in this case are confined to the planes parallel to the
equatorial plane. The components of Eq. (\ref{eq:modes}), which are
even in index 3, are:
\baray \label{mac_modes:2.1}
\lambda^2U_{3;3}&=& (\pi\rho G)^{-1}\delta\tilde\Pi - 2 A_3 U_{33}
-2 \Omega \tilde\beta''\lambda U_{3;3} \Comma\\
    \label{mac_modes:2.2}
\lambda^2U_{1;1}-2\Omega\lambda U_{2;1}
&=& (\pi G\rho)^{-1}\delta\tilde\Pi+(\Omega^2 - 2 A_1) U_{11}
- 2\Omega\tilde\beta\lambda U_{1;1}-2\Omega\tilde\beta'\lambda U_{2;1}\Comma\\
     \label{mac_modes:2.3}
\lambda^2U_{2;2}+2\Omega\lambda U_{1;2}
&=&  (\pi G\rho)^{-1}\delta\tilde\Pi+(\Omega^2 - 2 A_1) U_{22}
- 2 \Omega \tilde\beta \lambda U_{2;2} + 2\Omega\tilde\beta'\lambda U_{1;2}\Comma\\
     \label{mac_modes:2.4}
\lambda^2U_{1;2}-2\Omega\lambda U_{2;2}
&=& ( - 2 A_1 +\Omega^2) U_{12}
- 2 \Omega \tilde\beta \lambda U_{1;2} - 2\Omega\tilde\beta'\lambda U_{2;2}\Comma\\
     \label{mac_modes:2.5}
\lambda^2U_{2;1}+2\Omega\lambda U_{1;1}
&=& ( - 2 A_1 +\Omega^2) U_{21}
- 2 \Omega \tilde\beta \lambda U_{2;1} + 2\Omega\tilde\beta'\lambda U_{1;1}\Point
\earay
We add Eqs. (\ref{mac_modes:2.4}) and   (\ref{mac_modes:2.5})
and subtract Eqs. (\ref{mac_modes:2.2}) and   (\ref{mac_modes:2.3})
to find the following coupled equations for the toroidal modes
(note that $A_1 = A_2$ for Maclaurin spheroids)
\baray  \label{mac_modes:2.6}
\left(\lambda^2 + 2 \Omega \tilde\beta \lambda+4 A_1- 2 \Omega^2 \right)(U_{11}-U_{22})
-4 \Omega \lambda (1-\tilde\beta') U_{12} &=& 0\Comma \\\
     \label{mac_modes:2.7}
 \left(\lambda^2 + 2 \Omega \tilde\beta \lambda+4 A_1- 2 \Omega^2\right) U_{12}
+\Omega\lambda (1-\tilde\beta')(U_{11}-U_{22}) = 0 \Point
\earay
The characteristic equation for the toroidal modes is
\baray  \label{mac_modes:2.8}
&&\lambda^4 + 4 \tilde\beta \lambda^3\Omega+
   \lambda^2 (8 A_1 + 4 \tilde\beta^2\Omega^2 - 8\tilde\beta'\Omega^2 + 4\tilde\beta'^2\Omega^2)
   \nonumber \\
   &&\hspace{2cm}+
   \lambda (16 A_1 \tilde\beta\Omega - 8 \tilde\beta\Omega^3)
   +16 A_1^2 - 16 A_1\Omega^2 + 4\Omega^4
   = 0.
\earay
In the frictionless limit this can be written  as
\be
(\lambda^2 + 4 A_1 - 2 \Omega^2)^2 + 4 \Omega^2\lambda^2  =0\Comma
\ee
which is factorized by writing $\lambda=i\sigma$.
The two solutions are then
\be\label{eq:sig_tor}
\sigma_{1,2}=\Omega\pm\sqrt{4 A_1 - \Omega^2},
\ee
and there are
two complementary modes which are found by substituting $-\Omega$ for $\Omega$.
The modes are always real because $ 4A_1 > \Omega^2 $
for incompressible Maclaurin spheroids. This result
is in contrast to the modes of ordinary Maclaurin spheroids which
become dynamically unstable  at $4 B_{12} = \Omega^2$,  $\epsilon = 0.953$.
Our model of the superfluid Maclaurin spheroid also becomes
dynamically unstable at the same point, but only via
the toroidal modes derived from the perturbation equations for
$V_{ij}$, just as in EFE.

The real and imaginary parts of the dissipative toroidal modes are shown
in the Fig. 2, for the same values of $\eta$ as in
Fig. 1. The real parts of the modes
tend towards each other and merge in the  large friction limit. Note that
there are no neutral points associated with these modes and the necessary
condition for a point of bifurcation is not satisfied.
The damping of the modes is maximal, as in the case of the transverse shear
modes for $\eta = 1$, and decreases in both limits of $\eta\to 0$
and $\eta\to\infty$. In contrast to ordinary Maclaurin spheroids,
which become secularly unstable at the bifurcation point where
$2 B_{12} =\Omega^2$ and $\epsilon = 0.813$,
the new toroidal modes are stable at all values
of the eccentricity.
Our main conclusion is that the toroidal modes
associated with the relative motions of the superfluid and the normal
components always remain stable for incompressible Maclaurin spheroids.
In the case of the compressible Maclaurin spheroids the point of the
onset of secular instability may vary as a function of the adiabatic
index (in the case of a polytropic type of equation of state)
and hence the conclusions reached above should be verified for these
models separately.

\subsubsection{The relative pulsation mode}

To find the pulsation modes, which correspond to $m=0$,
we first add Eqs.
(\ref{mac_modes:2.2})-(\ref{mac_modes:2.3}) and subtract from the result
the Eq. (\ref{mac_modes:2.1}). In this manner we find that
\baray \label{mac_modes:3.1}
&&\left(\lambda^2/2+ \Omega\tilde\beta\lambda-\Omega^2+2 A_1\right)(U_{11}+U_{22})
       \nonumber \\
&&\hspace{2cm}+ 2 \Omega\lambda (1-\tilde\beta')(U_{1;2}-U_{2;1})
-(\lambda^2 + 4 A_3+2\Omega \tilde\beta'' \lambda) U_{33}=0.
\earay
Subtracting Eqs. (\ref{mac_modes:2.5}) and (\ref{mac_modes:2.4})
one finds
\baray
    \label{mac_modes:3.2}
&& \left(\lambda^2 + 2 \Omega\tilde\beta\lambda\right) (U_{1;2}-U_{2;1})
-\Omega\lambda(1-\tilde\beta') (U_{11}+U_{22}) = 0\Point
\earay
Equations (\ref{mac_modes:3.1}) and (\ref{mac_modes:3.2}) can be further
combined to a single equation:
\baray\label{mac_modes:3.3}
&&\left[\left(\lambda^2+2\Omega\tilde\beta\lambda - 2 \Omega^2 + 4 A_1\right)
(\lambda^2+2 \Omega\tilde\beta \lambda) + 4\Omega^2\lambda^2
(1-\tilde\beta')^2\right](U_{11}+U_{22})\nonumber \\
&&\hspace{2cm}-2\left[\left(\lambda^2+2\Omega\tilde\beta\lambda\right)
\left(\lambda^2+2\Omega\lambda\tilde\beta''+4A_3\right)\right]U_{33}=0\Point
\earay
The solution is found by supplementing these equations
by the  divergence free condition
\be \label{DIV}
\frac{U_{11}}{a_1^2}+\frac{U_{22}}{a_2^2}
+\frac{U_{33}}{a_3^2} = 0
\ee
or, in terms of the eccentricity 
$
\epsilon = 1 - a_3^2/a_1^2,
$
\be
(U_{11}+U_{22})(1-\epsilon^2)  + U_{33} = 0.
\ee
The third order characteristic equation is
\baray
&&(3 - 2\epsilon^2)\lambda^3 +
    (8\tilde\beta\Omega + 4\tilde\beta''\Omega
    - 4\tilde\beta\epsilon^2\Omega - 4\tilde\beta''\epsilon^2\Omega)\lambda^2+
   (4A_1 + 8A_3 - 8A_3\epsilon^2\nonumber \\
&&\hspace{2cm} + 2\Omega^2
   + 4\tilde\beta^2\Omega^2 - 8\tilde\beta'\Omega^2 +
      4\tilde\beta'^2\Omega^2 + 8\tilde\beta\tilde\beta''\Omega^2
      - 8\tilde\beta\tilde\beta''\epsilon^2\Omega^2) \lambda \nonumber \\
&&\hspace{2cm}+8A_1\tilde\beta\Omega + 16A_3\tilde\beta\Omega - 16A_3\tilde\beta
 \epsilon^2\Omega - 4\tilde\beta\Omega^3 = 0\Comma
\earay
where the trivial mode $\lambda = 0$ is neglected.
In the frictionless limit we find ($\lambda=i\sigma$ as before)
\be\label{eq:sig_pul_nondiss}
\sigma = \pm\left[
\frac{ 2 \Omega^2 + 4 A_1 + 8 A_3 (1-\epsilon^2)}{(3 - 2 \epsilon^2) }
\right]^{1/2} .
\ee
The pulsation modes for a sphere follow in the limit
$\epsilon, \Omega \to 0$:
for a sphere  $A_i/(\pi\rho G) = 2/3$, and Eq. (\ref{eq:sig_pul_nondiss})
reduces to $\sigma^2 = 8/3$  [$\sigma$ is given
in units of $(\pi\rho G)^{1/2}]$.
This result could be compared with the pulsation modes of an
ordinary sphere:  $\sigma^2 = 16/15$. Thus a superfluid sphere,
apart form the ordinary pulsations, shows pulsations
at frequencies roughly twice as large as the ordinary ones.

The real and imaginary parts of the dissipative pulsation modes
of a superfluid Maclaurin spheroid are shown
in the Fig. 3. The real parts of the modes
are weakly affected by mutual friction and closely resemble those of an
ordinary Maclaurin spheroid in the frictionless limit. These
are located, however, at higher frequencies. The symmetry
of the damping rate as a function of $\eta$ observed for the
transverse shear and toroidal modes is again observed in Fig. 3.
Note that the results
in Fig. 6 were obtained in the case $\beta''=0.$
The pulsation modes of the superfluid Maclaurin spheroid are
stable, as is the case for the ordinary Maclaurin spheroids.

\subsection{Modes of superfluid Jacobi ellipsoid}
The sequence of the Jacobi ellipsoids emerges from the Maclaurin
sequence at the bifurcation point $\epsilon= 0.813$
via a spontaneous breaking of symmetry in the plane
perpendicular to the rotation ($a_1\ne  a_2$).
The superfluid equilibrium figures are again identical to their
ordinary fluid counterparts and the defining relations
$a_1^2a_2^2A_{12} =a_3^2 A_3$ and $\Omega^2 = 2 B_{12}$
are unchanged.
Ordinary Jacobi ellipsoids are known to be stable against
second order harmonic perturbations while  they become
dynamically unstable against transformation into Poincar\` e's
pear shaped figures
through a mode belonging to third order harmonic perturbations.
If the sequence of Jacobi ellipsoids is parametrized in terms of
the variable ${\rm cos}^{-1}(a_3/a_1)$, it is stable between the
point of bifurcation from the Maclaurin sequence,
${\rm cos}^{-1}(a_3/a_1)=54.36$, and the point where 
Poincar\` e's figures bifurcate, ${\rm cos}^{-1}(a_3/a_1)=69.82$.
Here, by an explicit calculation, we verify that 
superfluid Jacobi ellipsoids do not develop new instabilities
via second order harmonic modes of the relative displacements.

\subsubsection{Relative odd modes}
The treatment of the oscillations of the Maclaurin spheroid of the
previous sections can be readily extended to the Jacobi ellipsoids
by lifting the degeneracy in indexes 1 and 2 and imposing
$A_{1}\ne A_{2}$.  The equations odd in index 3 are
\baray\label{jac_modes:4.10}
\lambda^2 U_{3;1} &=& -2 A_{3} U_{31}
-2 \Omega\tilde\beta''\lambda U_{3;1}\Comma\\
   \label{jac_modes:4.11}
\lambda^2 U_{3;2} &=& -2 A_{3} U_{32}
-2 \Omega\tilde\beta''\lambda U_{3;2}\Comma\\
   \label{jac_modes:4.12}
\lambda^2 U_{1;3} - 2 \Omega\lambda U_{2;3}
&=&\left(-2 A_{1}+\Omega^2\right)U_{13}
-2\Omega\tilde\beta\lambda U_{1;3}
- 2 \Omega\tilde\beta'\lambda U_{2;3}\Comma\\
   \label{jac_modes:4.13}
\lambda^2 U_{2;3} + 2 \Omega\lambda U_{1;3}
&=&\left(-2 A_{2}+\Omega^2\right)U_{23}
-2\Omega\tilde\beta\lambda U_{2;3}
+2 \Omega\tilde\beta'\lambda U_{1;3}\Point
\earay
Combining Eqs.
(\ref{jac_modes:4.10}) and (\ref{jac_modes:4.12}) and,
similarly, Eqs. (\ref{jac_modes:4.11}) and (\ref{jac_modes:4.13}),
after some manipulation we find
\baray
&&\left[\left(\lambda^2+2\Omega\tilde\beta''\lambda\right)
\left(\lambda^2+2\Omega\tilde\beta\lambda\right)+
2 \left(\lambda^2+2\Omega\tilde\beta\lambda\right) A_3
+ \left(\lambda^2+2\Omega\tilde\beta''\lambda\right) (2 A_1-\Omega^2)
\right] U_{13} \nonumber \\
&&\hspace{4cm} -2\Omega\lambda(1-\tilde\beta')
\left(\lambda^2+2\Omega\tilde\beta''\lambda + 2A_3\right) U_{23} = 0\Comma
\label{jac_modes:4.14}\\
&&\left[\left(\lambda^2+2\Omega\tilde\beta''\lambda\right)
\left(\lambda^2+2\Omega\tilde\beta\lambda\right)+
2 \left(\lambda^2+2\Omega\tilde\beta\lambda\right) A_3
+ \left(\lambda^2+2\Omega\tilde\beta''\lambda\right) (2 A_2-\Omega^2)
\right] U_{23} \nonumber \\
&&\hspace{4cm} + 2\Omega\lambda(1-\tilde\beta')
\left(\lambda^2+2\Omega\tilde\beta''\lambda + 2A_3\right)
 U_{13} = 0\Point
\label{jac_modes:4.15}
\earay
Equations (\ref{jac_modes:4.14}) and (\ref{jac_modes:4.15})  are
sufficient to determine the symmetric parts of the virials,
and any two of Eqs. (\ref{jac_modes:4.10})-(\ref{jac_modes:4.13})
can be used to find the antisymmetric parts.
The sixth order characteristic equation is
\baray \label{jac_modes:4.16}
&&\lambda^6 +  4 \Omega(\tilde\beta + \tilde\beta'') \lambda^5 +
    \Bigl[4 A_3 + 2\Omega(1-\tilde\beta')^2
   + 4\Omega^2 (\tilde\beta+\tilde\beta'')^2
  - (2A_1-\Omega^2) -(2A_2-\Omega^2) \Bigr]   \lambda^4
\nonumber\\
&&
   + \Bigl[16 A_3\tilde\beta\Omega + 8 A_3\tilde\beta''\Omega
   + 8\tilde\beta'' \Omega^2(1-\tilde\beta')^2
   + 16\tilde\beta^2\tilde\beta''\Omega^3  +
      16\tilde\beta\tilde\beta''^2\Omega^3 \nonumber\\
&&\hspace{3cm}
       - 2\tilde\beta\Omega (2A_1-\Omega^2) -  2\tilde\beta\Omega (2A_2-\Omega^2)
       - 4\tilde\beta''\Omega (2A_1-\Omega^2)
       - 4\tilde\beta''\Omega (2A_2-\Omega^2)\Bigr]  \lambda^3
\nonumber\\
&&
   + \Bigl[4 A_3^2 + 8 A_3 \Omega(1-\tilde\beta')^2
   + 16 A_3\tilde\beta^2\Omega^2 + 32 A_3\tilde\beta\tilde\beta''\Omega^2 +
      8\tilde\beta''^2 \Omega^3(1-\tilde\beta')^2
      + 16\tilde\beta^2\tilde\beta''^2\Omega^4
\nonumber\\
&&\hspace{3cm}
  -2 A_3 (2A_1-\Omega^2)- 8\tilde\beta\tilde\beta''\Omega^2
 (2A_1-\Omega^2) -4\tilde\beta''^2\Omega^2 (2A_1-\Omega^2)
   -2 A_3 (2A_2-\Omega^2)
\nonumber\\
&&\hspace{3cm}
  - 8\tilde\beta\tilde\beta''\Omega^2
 (2A_2-\Omega^2) -4\tilde\beta''^2\Omega^2 (2A_2-\Omega^2)
      + (2A_1-\Omega^2) (2A_2-\Omega^2) \Bigr]    \lambda^2
\nonumber\\
&&
+\Bigl[16 A_3^2\tilde\beta\Omega + 16 A_3\tilde\beta''
\Omega^2(1-\tilde\beta')^2
+ 32 A_3\tilde\beta^2\tilde\beta''\Omega^3
- 4 A_3\tilde\beta\Omega (2A_1-\Omega^2)
- 4 A_3\tilde\beta\Omega (2A_2-\Omega^2)\nonumber\\
&&\hspace{3cm} - 4 A_3\tilde\beta''\Omega (2A_1-\Omega^2)
      - 8\tilde\beta\tilde\beta''^2\Omega^3
      (2A_1-\Omega^2)
      - 4 A_3\tilde\beta''\Omega (2A_2-\Omega^2)
\nonumber\\
&&\hspace{3cm}
- 8\tilde\beta\tilde\beta''^2\Omega^3
      (2A_2-\Omega^2)
      + 4\tilde\beta''\Omega (2A_1-\Omega^2)
      (2A_2-\Omega^2) \Bigr]  \lambda   \nonumber  \\
&&
      +\Bigl[8 A_3^2 \Omega(1-\tilde\beta')^2
      + 16 A_3^2\tilde\beta^2\Omega^2
      - 8 A_3\tilde\beta\tilde\beta''\Omega^2 (2A_1-\Omega^2)
\nonumber\\
&& \hspace{3cm}
      - 8 A_3\tilde\beta\tilde\beta''\Omega^2 (2A_2-\Omega^2)
     + 4\tilde\beta''^2\Omega^2 (2A_1-\Omega^2)
      (2A_2-\Omega^2)\Bigr] = 0.
\earay
The real and imaginary parts of the dissipative odd parity
modes are shown in the Fig. 4. The
Jacobi sequence is parametrized in terms of ${\rm cos}^{-1} (a_3/a_1)$
starting off from the point of bifurcation of the Jacobi ellipsoid
from the Maclaurin sequence. The low frequency mode resembles
the rotational mode of the ellipsoid; its frequency decreases
with increasing friction. One of the  remaining two distinct high
frequency modes is almost unaffected by the dissipation, while
the other is suppressed close to the bifurcation point in a
monotonic manner. The damping rates of the odd modes are maximal
at $\eta = 1$ and tend to zero for both large and small friction.
The modes are damped
along the entire sequence; hence, we conclude that 
superfluid Jacobi ellipsoids
are stable against the odd second harmonic modes of oscillations.

\subsubsection{Relative even modes}
The explicit form of the even parity modes for the Jacobian sequence
is
\baray \label{jac_modes:4.21}
\lambda^2U_{3;3}&=& (\pi\rho G)^{-1}\delta\tilde\Pi -2A_3U_{33}
-2 \Omega \tilde\beta''\lambda U_{3;3}\Comma \\
\label{jac_modes:4.22}
\lambda^2U_{1;1}-2\Omega\lambda U_{2;1}
&=& (\pi G\rho)^{-1}\delta\tilde\Pi-2A_1U_{11}
+\Omega^2 U_{11}
- 2\Omega\tilde\beta\lambda U_{1;1}-2\Omega\tilde\beta'\lambda U_{2;1}
\Comma\\
\label{jac_modes:4.23}
\lambda^2U_{2;2}+2\Omega\lambda U_{1;2}
&=&  (\pi G\rho)^{-1}\delta\tilde\Pi-2A_2U_{22}
+\Omega^2 U_{22}
- 2 \Omega \tilde\beta \lambda U_{2;2} + 2\Omega\tilde\beta'\lambda U_{1;2}
\Comma\\
\label{jac_modes:4.24}
\lambda^2U_{1;2}-2\Omega\lambda U_{2;2}
&=& (\Omega^2 - 2A_1) U_{12}
- 2 \Omega \tilde\beta \lambda U_{1;2} - 2\Omega\tilde\beta'\lambda U_{2;2}
\Comma\\
\label{jac_modes:4.25}
\lambda^2U_{2;1}+2\Omega\lambda U_{1;1}
&=&  (\Omega^2 - 2A_2)U_{12}
- 2 \Omega \tilde\beta \lambda U_{2;1} + 2\Omega\tilde\beta'\lambda U_{1;1}\Point
\earay
These equations can be reduced to a simpler set of equations through
manipulations which eliminate the variations of the pressure tensor.
Explicitly, in the first step we subtract the Eqs. (\ref{jac_modes:4.22})
and (\ref{jac_modes:4.23}); in the second we sum
Eqs. (\ref{jac_modes:4.22}) and (\ref{jac_modes:4.23})
and subtract twice Eq. (\ref{jac_modes:4.21}). The result is
\baray
\label{jac_modes:4.26}
&&(\lambda^2/2+\Omega\tilde\beta\lambda - \Omega^2+ 2A_1) U_{11}
-(\lambda^2/2+\Omega\tilde\beta\lambda - \Omega^2+2A_2) U_{22}\nonumber\\
&&\hspace{2cm}
 - 2 \Omega\lambda (1-\tilde\beta') U_{12} = 0\Comma \\
\label{jac_modes:4.27}
&& (\lambda^2/2+\Omega\tilde\beta\lambda - \Omega^2+2A_1) U_{11}
+(\lambda^2/2+\Omega\tilde\beta\lambda - \Omega^2+ 2A_2) U_{22}\nonumber\\
&&\hspace{2cm}
-(\lambda^2+2\Omega\tilde\beta''\lambda+4A_3)U_{33}
+2\Omega\lambda(1-\tilde\beta')(U_{1;2}-U_{2;1})
=0\Point
\earay
Further we add and subtract Eqs.
 (\ref{jac_modes:4.24}) and (\ref{jac_modes:4.25}) to find
\baray\label{4.28}
\left[\lambda^2+\Omega\tilde\beta\lambda-4B_{12} + 2(A_1+A_2)\right] U_{12}
+\Omega(1-\tilde\beta')\lambda(U_{11}-U_{22})=0\Comma\\
\label{jac_modes:4.29}
(\lambda^2+2\Omega\tilde\beta\lambda) (U_{1;2}-U_{2;1}) -
 \Omega(1-\tilde\beta')\lambda (U_{11}+U_{22}) +2 (A_1 -A_2)U_{12} = 0\Point
\earay
Equations (\ref{jac_modes:4.26})-(\ref{jac_modes:4.29}), supplemented by the
divergence free condition, Eq. (\ref{DIV}), are sufficient to
determine the modes. The characteristic equation is of
seventh order, excluding the trivial root $\lambda =0$;
in the frictionless limit the characteristic equation is of 
third order.  The explicit form of these equations is cumbersome
and will not be given here.

The real and imaginary parts of the dissipative even parity
modes are shown in Fig. 5. For each
member of the sequence, the eigenvalues
of the two high frequency modes are suppressed
and that of the low-frequency mode is amplified as
the dissipation increases. As in the case of the odd modes
the damping rates of the even parity modes are maximal
at $\eta = 1$ and tend to zero for both large and small
friction. The damping rates are again positive
along the entire sequence and we conclude that 
superfluid Jacobi ellipsoids
are stable against the even parity second harmonic modes
of oscillations.

\subsection{Modes of superfluid Roche ellipsoid}
In this section we extend the previous discussion of isolated
ellipsoids to binary star systems, and consider the simplest
case -- the Roche problem. The classical Roche binary
consists of a finite size ellipsoid (primary of mass $M$)
and a point mass (secondary of mass $M'$) rotating about
their common center of mass with an angular velocity $\Omega$.
The new ingredient in the problem of the equilibrium and stability
of the primary is the tidal Newtonian gravitational field
of the secondary. Place the center of the
coordinate system at the center of mass of the primary with the
$x_1$-axis pointing to the center of mass of the secondary
and $x_3$-axis along the vector ${\bf \Omega}$.
The equation of motion for a fluid element of the primary
in the frame rotating with angular velocity
$\Omega$ is, then (EFE, Chap. 8,  Sec. 55)
\be
\rho_\alpha D_\alpha u_{\alpha,i}=-{\partial p_\alpha\over
\partial x_i}-\rho_\alpha{\partial(\phi+\phi')\over\partial x_i}
+{1\over 2}\rho_\alpha{\partial\vert\Omvec\crossprod\xvec\vert^2
\over\partial x_i}+2\rho_\alpha\epsilon_{ilm}u_{\alpha,l}\Omega_m
+F_{\alpha\beta,i},
\label{eq:eulerroche}
\ee
where the tidal potential of the secondary,
up to quadratic terms in $x_i/R$, is
\be
\phi' = \frac{GM'}{R}
\left(1+\frac{x_1}{R}+\frac{2x_1^2-x_2^2-x_3^2}{2R^2}\right).
\ee
The modified Keplerian rotation frequency for circular
orbits, which is consistent with the first order virial equations,
is\cite{LRS}
\be\label{KEPLER}
\Omega^2 = (1+P)\phi_0 (1+\delta),
\ee
where $P = M/M'$ is the mass ratio,
$\phi_0 = GM'/R^3$ is the tidal potential at the
origin of the primary, and
$\delta$ is the quadrupole part of the
tidal field. The latter correction
to the Keplerian frequency is maximal at
the Roche limit where $\delta \sim 0.13$ \cite{LRS}.
For the sake of simplicity this correction is
dropped in the following, as it does
not enter into the
analysis of the
stability of the Roche ellipsoid
for displacements that are linear functions of the
coordinates. However, the
relation between the frequency $\Omega$ and
the orbital separation is now determined within an
accuracy $\delta\ll 1$.
As in the case of the solitary ellipsoids, we find that
the equilibrium figure of the superfluid Roche ellipsoid
is identical to its ordinary fluid counterpart.

To treat  Roche ellipsoids we modify Eq. (\ref{eq:modes}) to
\baray
\lambda^2 U_{i;j}-2\epsilon_{il3}\Omega_3 \lambda U_{l;j} &=&
\delta_{ij}(\pi\rho G)^{-1}\delta\tilde\Pi-A_{i}U_{ij}
-2\Omega\lambda\tilde\beta_{ik}U_{k;j}\nonumber\\
& +& (\Omega^2-\phi_0) U_{ij} -
\Omega^2\delta_{i3}U_{3j} + 3 \phi_0 \delta_{i1} U_{1j}\Comma
\label{eq:modes_roche}
\earay
where all frequencies are measured in units of $(\pi \rho G)^{1/2}$.
Equation (\ref{eq:modes_roche}) is appropriate for finding the second harmonic
modes of oscillations of Roche ellipsoids.

\subsubsection{Relative odd modes}
The equations determining the modes even and odd in index 3
form separate sets. We start with the
modes belonging to $l=2 $ and $m = -1, 1 $ displacements,
which are odd in index 3; for these,
\baray\label{ro_modes:5.10}
\lambda^2 U_{3;1} &=& -(2 A_{3}+\phi_0) U_{31}
-2 \Omega\tilde\beta''\lambda U_{3;1}\Comma\\
   \label{ro_modes:5.11}
\lambda^2 U_{3;2} &=& -2 (A_{3}+\phi_0) U_{32}
-2 \Omega\tilde\beta''\lambda U_{3;2}\Comma\\
   \label{ro_modes:5.12}
\lambda^2 U_{1;3} - 2 \Omega\lambda U_{2;3}
&=&\left(-2 A_{1}+\Omega^2 + 2\phi_0\right)U_{13}
-2\Omega\tilde\beta\lambda U_{1;3}
- 2 \Omega\tilde\beta'\lambda U_{2;3}\Comma\\
   \label{ro_modes:5.13}
\lambda^2 U_{2;3} + 2 \Omega\lambda U_{1;3}
&=&\left(-2 A_{2}+\Omega^2 -\phi_0\right)U_{23}
-2\Omega\tilde\beta\lambda U_{2;3}
+2 \Omega\tilde\beta'\lambda U_{1;3}\Point
\earay
On combining Eqs.
(\ref{ro_modes:5.10}) and (\ref{ro_modes:5.12}) and,
similarly, Eqs. (\ref{ro_modes:5.11}) and (\ref{ro_modes:5.13})
we obtain
\baray\label{ro_modes:5.14}
&&\left[(\lambda^2+2\Omega\tilde\beta''\lambda+2A_3+\phi_0)
 (\lambda^2+2\Omega\tilde\beta\lambda)
+ (\lambda^2+2\Omega\tilde\beta''\lambda) (2 A_1-\Omega^2-2\phi_0) \right] U_{13}\nonumber \\
&&\hspace{2cm}-2\Omega(1-\tilde\beta')\lambda\left[\lambda^2+2\Omega\tilde\beta''\lambda
+ 2 A_3+\phi_0\right] U_{32} = 0\Comma\\
\label{ro_modes:5.15}
&&\left[(\lambda^2+2\Omega\tilde\beta''\lambda+2A_3+\phi_0)
(\lambda^2+2\Omega\tilde\beta\lambda)
+ (\lambda^2+2\Omega\tilde\beta''\lambda) (2 A_2-\Omega^2+\phi_0) \right]
U_{23}\nonumber \\
&&\hspace{2cm}+2\Omega(1-\tilde\beta')\lambda
\left[\lambda^2+2\Omega\tilde\beta''\lambda
+ 2 A_3+\phi_0\right] U_{13} = 0\Point
\earay
The $U_{ij}$ are symmetric under interchange of their indexes,
and Eqs.
(\ref{ro_modes:5.14}) and (\ref{ro_modes:5.15}) completely determine
the modes [any two of Eqs. [\ref{ro_modes:5.10}]-[\ref{ro_modes:5.13}] may
be used to find the antisymmetric parts of $U_{i;j}$].
The real and imaginary parts of the dissipative odd parity
modes are shown in Fig. 6, for the
case of an equal mass binary ($P=1$). The relative modes of Roche ellipsoids
for other values of the mass ratio display behavior similar to the $P=1$
case. The
Roche sequence is parametrized in terms of ${\rm cos}^{-1} (a_3/a_1)$.
The low frequency mode resembles
the rotational mode of the ellipsoid; its frequency decreases
with increasing friction.  The high frequency modes tend towards
each other and merge in the limit of slow rotation; in
the opposite limit the modes remain unaffected by the dissipation.
There are three distinct rates for the damping of oscillations.
These are maximal at $\eta = 1$ and tend to zero in both limits
of large and small friction, as was the case for the Maclaurin
and Jacobi ellipsoids.
The damping rates are positive
along the entire sequence. We conclude that 
superfluid Roche ellipsoids do not develop instabilities
via the second order harmonic odd modes of relative oscillation.

\subsubsection{Relative Even Modes}
As is well known, Roche ellipsoids develop a dynamical instability
via the second order even parity modes beyond the Roche limit, which
is the point of closest approach of the primary to the secondary.
If  viscous dissipation is allowed for,  Roche ellipsoids become
secularly unstable at the Roche limit via an even parity mode and
before dynamical instability sets in. We have seen that 
Maclaurin spheroids do not develop any instabilities (i.e. neither
dynamical nor secular) via the modes associated with the
$U_{ij}$  in the presence of superfluid dissipation.
The extension of the theory above
to superfluid Roche ellipsoids, as we
show now, does not reveal any new instabilities
in the presence of superfluid dissipation, again in contrast to
the analysis  based on the ordinary viscous dissipation.

The explicit form of the even parity modes for the Roche sequence
is
\baray
\label{ro_modes:5.16}
\lambda^2U_{3;3}&=& (\pi\rho G)^{-1}\delta\tilde\Pi -(2A_3+\phi_0)U_{33}
-2 \Omega \tilde\beta''\lambda U_{3;3}\Comma \\
\label{ro_modes:5.17}
\lambda^2U_{1;1}-2\Omega\lambda U_{2;1}
&=& (\pi G\rho)^{-1}\delta\tilde\Pi-(2A_1-\Omega^2-2\phi_0)U_{11}
- 2\Omega\tilde\beta\lambda U_{1;1}-2\Omega\tilde\beta'\lambda U_{2;1}
\Comma\\
\label{ro_modes:5.18}
\lambda^2U_{2;2}+2\Omega\lambda U_{1;2}
&=&  (\pi G\rho)^{-1}\delta\tilde\Pi-(2A_2-\Omega^2+\phi_0)U_{22}
+\Omega^2 U_{22}
- 2 \Omega \tilde\beta \lambda U_{2;2} + 2\Omega\tilde\beta'\lambda U_{1;2}
\Comma\\
\label{ro_modes:5.19}
\lambda^2U_{1;2}-2\Omega\lambda U_{2;2}
&=& (\Omega^2 +2\mu- 2A_1) U_{12}
- 2 \Omega \tilde\beta \lambda U_{1;2} - 2\Omega\tilde\beta'\lambda U_{2;2}
\Comma\\
\label{ro_modes:5.20}
\lambda^2U_{2;1}+2\Omega\lambda U_{1;1}
&=&  (\Omega^2-\phi_0 - 2A_2)U_{12}
- 2 \Omega \tilde\beta \lambda U_{2;1} + 2\Omega\tilde\beta'\lambda U_{1;1}\Point
\earay
These equations can be reduced to a simpler set of equations through
manipulations which eliminate variations of the pressure tensor.
Using the symmetry properties of the virials we
first subtract Eqs. (\ref{ro_modes:5.17})
and (\ref{ro_modes:5.18}),  then  sum
Eqs. (\ref{ro_modes:5.17}) and (\ref{ro_modes:5.18})
and subtract twice  Eq. (\ref{ro_modes:5.16}) to obtain
\baray
\label{ro_modes:5.6}
&&(\lambda^2/2+\Omega\tilde\beta\lambda - \Omega^2-2\phi_0+2A_1) U_{11}
-(\lambda^2/2+\Omega\tilde\beta\lambda - \Omega^2+\phi_0+2A_2) U_{22}\nonumber\\
&&\hspace{2cm}- 2 \Omega\lambda (1-\tilde\beta') U_{12} = 0 \Comma\\
\label{ro_modes:5.7}
&& (\lambda^2/2+\Omega\tilde\beta\lambda - \Omega^2
- 2 \phi_0+ 2 A_1) U_{11}
+(\lambda^2/2+\Omega\tilde\beta\lambda
- \Omega^2+\phi_0+2A_2) U_{22}\nonumber\\
&&\hspace{2cm}
-(\lambda^2+2\Omega\tilde\beta''\lambda+2\phi_0+4A_3)U_{33}
+2\Omega\lambda(1-\tilde\beta')(U_{1;2}-U_{2;1})
=0 \Point
\earay
Further we add and subtract Eqs. (\ref{ro_modes:5.19}) and
(\ref{ro_modes:5.20})  to find
\baray\label{ro_modes:5.21}
\left[\lambda^2+2\Omega\tilde\beta\lambda+2(A_1+A_2)-2\Omega^2-\phi_0\right] U_{12}
+\Omega(1-\tilde\beta')\lambda(U_{11}-U_{22})=0\Comma\\
\label{ro_modes:5.22}
(\lambda^2+2\Omega\tilde\beta\lambda) (U_{1;2}-U_{2;1})
- \Omega(1-\tilde\beta')\lambda (U_{11}+U_{22})-[3 \phi_0 -2(A_1-A_2)] U_{12} = 0\Point
\earay
Equations (\ref{ro_modes:5.21})-(\ref{ro_modes:5.22}), supplemented by the
divergence free condition, Eq. (\ref{DIV}), are sufficient to
determine the unknown virials.

The real and imaginary parts of the dissipative even parity
modes are shown in Fig. 7.
For each member of the sequence the eigenvalues of
the two high frequency modes are suppressed and that
of the low frequency one is amplified with increasing dissipation.
In effect
these modes merge in the slow rotation limit. The modes do not
become neutral at any point along the frictionless sequence and, hence, the
necessary condition for the onset of dynamical instability
is not achieved. Note that our model for the Roche ellipsoids is
dynamically unstable as is its classical counterpart, but via the modes
governed by the Eq. (\ref{meanflow_basic}) modified appropriately to
include the external tidal potential.  We do not repeat the
mode analysis for the virials $V_{ij}$ as it is a complete
analogue of the analysis in EFE.  As in the case of odd modes
the damping rates of even parity modes are maximal
at $\eta = 1$ and tend to zero in both limits of large and small
friction. The damping rates are positive along the entire sequence
and we conclude that superfluid Roche ellipsoids are secularly
stable against the even parity second harmonic modes of oscillations
associated with the relative motions between the superfluid and normal
components.

\section{Viscosity and Gravitational Radiation}
\label{viscous}

Above, we neglected viscous dissipation in computing normal modes of a
normal fluid-superfluid mixture. However, for a single fluid, viscous
dissipation is important for understanding stability, for it is
responsible for secular instability. Although viscous terms spoil
the calculation of the modes of uniform ellipsoids from moment equations
formally, when the dissipative time-scale is long, one can include
them perturbatively (e.g. EFE, Chap. 5, \S37b).

The inclusion of viscosity is slightly more complicated for a mixture
of superfluid and normal fluid because viscous dissipation only
operates on the normal fluid. The separation of  Eq. (\ref{alphapert})
for the fluid displacements into Eqs. (\ref{eq:common}) and
(\ref{eq:different}) for the common and differential fluid displacements,
$\xivec_\pm$, is possible because the only form of dissipation,
the mutual friction force, included in Eq. (\ref{alphapert})
only depends on $d\xivec_-/dt$. Viscous dissipation depends on
$\xivec_N$ only, and, in a formal sense, the dynamics no longer
separate into the independent dynamics of $\xivec_\pm$.

If we assume that the time-scale associated with viscous dissipation
is relatively long, then we can include it perturbatively. The
calculation is a bit more subtle than for a single fluid, because
we have to deduce $\xivec_N$ for the modes. This brings up an issue
that we glossed over earlier, in setting up the calculation of
modes in \S\ref{ve2pe}: even when computing the modes that arise
from the equation for $\xivec_-$ alone, the equation for
$\xivec_+$ must be satisfied, and viceversa. The simplest way
for this to work is for $\xivec_+$ to vanish when $\xivec_-$
is nonzero and vice versa. In fact it is easy to show that
this is a reasonable solution provided that the Eulerian pressure
perturbations are $\delta p_\alpha=-\xi_{\alpha,l}\partial p_\alpha
/\partial x_l$, a situation that arises naturally for adiabatic
perturbations, where the Lagrangian pressure perturbations are
$\Delta_\alpha p_\alpha=-\Gamma_\alpha\partial\xi_{\alpha,l}/
\partial x_l$, {\it and} the perturbations are solenoidal,
so that $\partial\xi_{\alpha,l}/\partial x_l=0$, as is true
for all modes considered in this paper.
\footnote{More realistically, one would also need to consider
non-adiabatic effects, such as might arise from perturbations
from $\beta-$equilibrium; see e.g. Lindblom and Mendell
\cite{LINDBLOM}. These would tend to couple $\xivec_\pm$, but if small,
could be computed perturbatively, as we do here for viscosity, which
also couples $\xivec_\pm$.}
To see how this works,
consider a mode of Eq. (\ref{eq:different}), and examine under
what conditions Eq. (\ref{eq:common}) will be satisfied. Then,
using
\be
\delta p_\alpha=-\xi_{\alpha,l}{\partial p_\alpha\over
\partial x_l}=-f_\alpha\xi_{\alpha,l}{\partial p\over
\partial x_l}
\label{palphapert}
\ee
and [substituting the definition of
$\xivec_+$, and $\rho_\alpha=f_\alpha\rho$
into Eq. (\ref{eq:eulerphi})]
\be
\delta\phi\equiv -G\int_V{d^3\xp
\rho(\xpvec)\xi_{+,l}(\xpvec)
{\partial\over\partial\xp_l}\biggl({1\over
\vert\xvec-\xpvec\vert}\biggr)}
\ee
it is easy to see that Eq. (\ref{eq:common})
only depends on $\xivec_+$. But since the normal modes of
Eq. (\ref{eq:different}) have different frequencies than
normal modes of Eq. (\ref{eq:common}), we must have
$\xivec_+=0$ when $\xivec_-\neq 0$.
Since, in general,
\be
\xivec_S=\xivec_++f_N\xivec_-\qquad
\xivec_N=\xivec_+-f_S\xivec_-,
\ee
we conclude that, for modes of Eq. (\ref{eq:different}),
$\xivec_S=f_N\xivec_-$ and $\xivec_N=-f_S\xivec_-$, when Eulerian
pressure perturbations are given by Eq. (\ref{palphapert}).
Similarly, since Eq. (\ref{eq:different}) only depends on
$\xivec_-$, for modes with $\xivec_+\neq 0$, we must
have $\xivec_-=0$ and, therefore, $\xivec_S=\xivec_N=\xivec_+$,
assuming Eq. (\ref{palphapert}). In particular, if the kinematic 
viscosity $\nu$ is held constant [see EFE, Chap. 5, Sec. 36, Eq. (111)], 
for modes with $\xivec_+\neq 0$, the viscous dissipation rate 
is smaller by a factor of $f_N$ than it is for a single fluid 
with the same background and displacement $\xivec_+$, as might 
have been expected qualitatively (i.e. for small normal fluid 
density, the viscous dissipation must be diminished).
For perturbations with $\xivec_-\neq 0$ and displacements
that are linear functions of the coordinates, we must add
\be
-5f_S\nu\biggl({1\over a_j^2}{dU_{i;j}\over dt}
+{1\over a_i^2}{dU_{j;i}\over dt}\biggr)
\ee
to the right hand
side of Eq. (\ref{eq:differentmom1}) to include the effects
of viscous dissipation.

Perturbations with $\xivec_-\neq 0$ emit {\it no} gravitational
radiation because $\xivec_+=0$ for them, and therefore there
are no perturbations of the quadrupole moment or any other net
mass currents associated with them. Gravitational radiation
is emitted by the modes in which the two fluids move together
at the same rate as for a single fluid (e.g. ref. \cite{GRAVRAD}).
Thus, none of the new modes of a superfluid-normal fluid mixture
found here is affected by gravitational radiation at all. 

\section{Conclusions}

Despite a number of simplifying assumptions, the study of the oscillation
modes of uniform ellipsoids is useful for understanding the equilibrium
and stability of real neutron stars, at least qualitatively. Moreover, the
theory is simple enough that it can be extended readily to include modifications
to the underlying physics; here, we have considered new features that arise
because a neutron star contains a mixture of normal fluid and superfluid
coupled by mutual friction. The theory of uniform ellipsoids is interesting
from the viewpoint of mathematical physics because it is solvable exactly,
and it may also have applications to other physical systems, such as the
physics of trapped, rotating Bose-Einstein condensates.

In this paper  we have extended  previous treatments of the
oscillation modes of ellipsoidal figures of equilibrium to
the case of a mixture of normal fluid and superfluid.
The basic equations of motion for the
two fluid hydrodynamics include mutual friction exactly,
since the frictional forces depend linearly
on the relative velocity between
the two fluids, and vanish in the background where the two
fluids move together. In addition the fluids are coupled via
mutual gravitational attraction, which is also treated without
further approximations. As a result our relations closely
resemble Chandrasekhar's tensor virial equations, even though
they are intrinsically dissipative due to the mutual friction.
While we have developed these moment equations for general
underlying equilibria, they are most useful for perturbations
around uniform backgrounds, in which case the moment equations
of various orders decouple and yield exact solutions for the
normal modes.

Quite generally, there are two classes of modes for small
perturbations, one class
in which the two fluids move together and the other in which there
is relative motion between them. The former are identical to the
modes found for a single fluid. As a result our models of superfluid
Maclaurin and Roche ellipsoids  undergo dynamical instabilities
with respect to these modes  which are
indistinguishable from what is found for their classical counterparts.
When ordinary viscous dissipation is included they are also subject
to secular instabilities related to the modes in which two fluids move
together. If the kinematic viscosity is held constant, then 
the rate of viscous dissipation, when computed in the  ``low Reynolds
number'' approximation, is diminished by a factor $f_N$, the 
fraction of the total mass in the normal fluid (see however below).

The modes involving the relative 
motion between the fluids are completely new and
are shown to be  stable along the entire sequences of the incompressible
Maclaurin, Jacobi and Roche ellipsoids independent of the
magnitude of the phenomenological mutual friction. These modes also
do not become neutral at selected  points along any sequence and the
necessary condition for the point of bifurcation is not achieved.
Our main conclusion is that mutual friction does not drive
secular instabilities in  incompressible and irrotational ellipsoids.
In addition we find that even though the new modes are subject
to viscous dissipation (a consequence of viscosity of the normal matter),
they do not emit gravitational radiation, and are therefore immune to
any instabilities associated with gravitational radiation, irrespective
of their modal frequencies.

The results summarized above hold within a combined 
framework of two-fluid superfluid hydrodynamics,  
Newtonian gravity,  and the ellipsoidal approximation, as 
formulated in EFE. Each of these elements of our approach contains
a number of 
simplifying approximations which need to be relaxed in realistic 
applications to neutron stars. For example, to treat the mixture 
of neutron and proton superfluids in the neutron star cores, 
the one-constituent two-fluid superfluid hydrodynamics must be replaced
by the hydrodynamics of the  multi-constituent superfluid mixtures, 
in which  case the mutual entrainment of the superfluids and  
deviations from $\beta$-equilibrium must be accounted for 
(see Ref. \cite{LINDBLOM} for a discussion of these effects and 
their impact on neutron star oscillations within the real 
energy functional method). 
We anticipate that these effects can be 
incorporated in the tensor virial approach 
in a perturbative manner and the results of previous
sections will hold in leading order of the perturbation expansion.
On the other hand, relaxing the incompressible approximation and, hence, 
including the partial pressures of the superfluid and the normal fluid, 
will lead to non-perturbative effects as the pressure terms significantly 
alter the balance between gravitational attraction and  centrifugal 
stretching. As is well known, the points of the onset of the dynamical 
and secular instabilities of compressible ellipsoids 
depend on the adiabatic index of the underlying 
polytropic equation of state. As noted in Sec. 3, the conclusions reached 
with respect to the stability of the incompressible ellipsoids should 
be verified for the compressible models anew. 
The differences in the pressures (or equations of states) 
of the normal and superfluid 
phases, however, are typically small in neutron stars, 
since the condensation energy is negligible compared to the 
degeneracy pressures of interacting 
Fermi-liquids. The coupling between the partial pressures of the noraml 
fluid and superfluid, therefore, can be treated perturbatively.

Another effect that needs to be included is the density 
dependence of the mutual friction
coefficients and the kinematic viscosity. For example, 
in neutron stars the kinematic viscosity is density dependent
in general, explicitly as a result of  to the density dependence of the
phase space of normal quasiparticles undergoing collisions, implicitly
because of the density dependence of the in-medium scattering amplitudes 
(see for further details Ref. \cite{CUTLER_LINDBLOM}).
The ramification
for comparison of the secular instabilities of the normal fluid 
and superfluid ellipsoids is that the modifications of the 
rate of the viscous dissipation will depend in a non-trivial 
manner on the fraction of the normal fluid in the system. 

The entrainment, $\beta$-nonequilibrium, compressibility, e. t. c. 
will couple the relative and center-of-mass modes in general. 
Such a ``mixing'', as discussed in Sec. 5, implies  $\xivec_+\neq 0$ 
for the relative modes and, similarly, $\xivec_-\neq 0$ for the 
center-of-mass modes. Therefore, the mutual friction might tend to drive the
center-of-mass modes secularly unstable; if they
emit gravitational radiation, the mutual friction will
suppress the gravitational radiation induced instabilities.
One of the important issues to be addressed by the
future work is the magnitude of the ``mixing'' of the
modes for  general equilibria and the corresponding times
scales.

\acknowledgements
A.S. acknowledges  the Nederlandse Organisatie
voor Wetenschappelijk Onderzoek for its support at KVI Groningen
via  the Stichting voor
Fundamenteel Onderzoek der Materie,
the Institute for Nuclear Theory at the
University of Washington for its hospitality and the
Department of Energy for partial support. I.W. acknowledges
partial support for this project from NASA.

\appendix
\section{``Virial'' Equations and Perturbations}

The two fluids need not occupy the same volume, and we shall suppose
that fluid $\alpha$ occupies a volume $V_\alpha$.
Taking the zeroth moment of Eq. (\ref{eq:euler})
amounts to integrating over $V_\alpha$; doing so, we obtain\footnote{We
assume that $p_\alpha=0$ on the bounding surface of $V_\alpha$.} the
``first order `virial' equation''\footnote{We put the word ``virial''
in quotes because the equations are dissipative.}
\baray
{d\over dt}\biggl(\int_{V_\alpha}{d^3x\rho_\alpha u_{\alpha,i}}
\biggr)&=&2\epsilon_{ilm}\Omega_m\int_{V_\alpha}{d^3x\rho_\alpha
u_{\alpha,l}}+(\Omega^2\delta_{ij}-\Omega_i\Omega_j)
\int_{V_\alpha}{d^3x\rho_\alpha x_j}
\nonumber\\& &
-(1-\delta_{\alpha\beta})
\int_{V_\alpha}{d^3x\rho_\alpha{\partial\phi_\beta\over\partial x_i}}
+\int_{V_\alpha}{d^3xF_{\alpha\beta,i}}.
\label{eq:v1}
\earay
Apart from inertial forces, which do not couple the two fluids,
there are two forces that do couple them: gravity and friction.
The net, mutual gravitational force between the fluids only vanishes
if they (i) occupy the same volume and (ii) have densities
that are proportional to one another (i.e. $\rho_S\propto
\rho_N$). The mutual friction force is nonzero as long as the fluids
move relative to one another. Thus we see that, for a two fluid
mixture, the zeroth moment of Eq. (\ref{eq:euler}) is not trivial,
as it would be for a single fluid (as in EFE). Note, though, that
the center of mass motion of the combined system is trivial; i.e.,
if the center of mass of the combined system starts out at
$\xvec=0$ with zero velocity, it does not move.\footnote{There is
a slight subtlety that has not been stated explicitly in deriving
Eq. (\ref{eq:v1}). The mutual friction force is nonzero only in
the {\it overlap} volume of the two fluids. This restriction is
necessary to derive conservation of total momentum for the combined
fluids. We might be interested in situations involving long range
coupling, such as between the core superfluid and crustal normal fluid.
For such a coupling to occur, we need to introduce long range fields
(e.g. magnetic fields) capable of transmitting forces between
fluid elements at different points in space.}

Taking the first moment of Eq. (\ref{eq:euler}) results in the
second order ``virial'' equation
\baray
{d\over dt}\biggl(\int_{V_\alpha}{d^3x\rho_\alpha x_ju_{\alpha,i}}
\biggr)&=&2\epsilon_{ilm}\Omega_m\biggl(\int_{V_\alpha}
{d^3x\rho_\alpha x_ju_{\alpha,l}}\biggr)
+\Omega^2I_{\alpha,ij}-\Omega_i\Omega_kI_{\alpha,kj}
\nonumber\\& &
+2{\cal T}_{\alpha,ij}+\delta_{ij}\Pi_{\alpha}
+{\cal M}_{\alpha,ij}+(1-\delta_{\alpha\beta}){\cal M}_{\alpha\beta,ij}
+{\cal F}_{\alpha\beta,ij},
\label{eq:v2}
\earay
where
\baray
I_{\alpha,ij}&\equiv&\int_{V_\alpha}{d^3x\,\rho_\alpha x_ix_j}\nonumber\\
\Pi_\alpha&\equiv&\int_{V_\alpha}{d^3x\,p_\alpha}\nonumber\\
{\cal T}_{\alpha,ij}&\equiv&{1\over 2}\int_{V_\alpha}{d^3x\rho_\alpha
u_{\alpha,i}u_{\alpha,j}}\nonumber\\
{\cal M}_{\alpha,ij}&\equiv&
-{G\over 2}\int_{V_\alpha}{d^3x\,d^3\xp\rho_\alpha(\xvec)\rho_\alpha(\xpvec)
(x_i-\xp_i)(x_j-\xp_j)\over\vert\xvec-\xpvec\vert^3}\nonumber\\
{\cal M}_{\alpha\beta,ij}&\equiv&
-G\int_{V_\alpha}{d^3x}\int_{V_\beta}{d^3\xp\rho_\alpha(\xvec)
\rho_\beta(\xpvec)x_j(x_i-\xp_i)\over\vert\xvec-\xpvec\vert^3}
\nonumber\\
{\cal F}_{\alpha\beta,ij}&\equiv&\int_{V_\alpha}{d^3x\,x_j
F_{\alpha\beta,i}}.
\earay
When there is only one fluid present, this equation reduces to the
results found in Chap. 2 of EFE. There are two new terms here:
there is a term that arises from
the mutual gravitational forces of 
the two fluids (${\cal M}_{\alpha\beta,ij}$),
and also a term from the mutual friction (${\cal F}_{\alpha\beta,ij}$).

Consider first the variation of the first order 
virial equation under the influence
of perturbations. Most terms are simple to compute, but we must take special
care in computing
\be
-\delta\int_{V_\alpha}{d^3x\rho_\alpha{\partial\phi_\beta
\over\partial x_i}}=
-\delta G\int_{V_\alpha}{d^3x}\int_{V_\beta}{d^3\xp
\rho_\alpha(\xvec)\rho_\beta(\xpvec)(x_i-\xp_i)\over\vert\xvec-\xpvec\vert^3}.
\ee
In computing the necessary variation, think of $\xvec$ as having a label
$\alpha$ and $\xpvec$ as having a label $\beta$. It is then easy to find
that
\baray
-\delta \int_{V_\alpha}{d^3x\rho_\alpha{\partial\phi_\beta\over\partial x_i}}
=-G\int_{V_\alpha}{d^3x\rho_\alpha(\xvec)
\xi_{\alpha,l}(\xvec){\partial\over\partial x_l}}
\int_{V_\beta}{d^3\xp\rho_\beta(\xpvec)(x_i-\xp_i)
\over\vert\xvec-\xpvec\vert^3}
\nonumber\\
+G\int_{V_\beta}{d^3x\rho_\beta(\xvec)
\xi_{\beta,l}(\xvec){\partial\over\partial x_l}}
\int_{V_\alpha}{d^3\xp\rho_\alpha(\xpvec)
(x_i-\xp_i)\over\vert\xvec-\xpvec\vert^3}
\label{gravab}
\earay
which is manifestly antisymmetric on $\alpha\leftrightarrow\beta$.
Assuming $V_\alpha=V_\beta=V$ and $\rho_\alpha=f_\alpha\rho(\xvec)$ in the background
equilibrium, we can simplify this to
\be
-\delta\int_{V_\alpha}{d^3x\rho_\alpha{\partial\phi_\beta
\over\partial x_i}}=Gf_\alpha f_\beta
\int_V{d^3x\rho(\xvec)[\xi_{\beta,l}(\xvec)-\xi_{\alpha,l}(\xvec)]{\partial\over\partial x_l}}
\int_V{d^3\xp\rho(\xpvec)(x_i-\xp_i)\over\vert\xvec-\xpvec\vert^3}.
\ee
Gathering terms, we find that the perturbed first order virial equation is
\baray
{d^2\over dt^2}\biggl(f_\alpha\int_V{d^3x\rho\xi_{\alpha,i}}\biggr)
&=& 2\epsilon_{ilm}\Omega_m{d\over dt}\biggl(f_\alpha\int_V{d^3x\rho\xi_{\alpha,l}}\biggr)
+(\Omega^2\delta_{ij}-\Omega_i\Omega_j)f_\alpha\int_V{d^3x\rho\xi_{\alpha,j}}
\nonumber\\
& &+Gf_\alpha f_\beta
\int_V{d^3x\rho(\xvec)[\xi_{\beta,l}(\xvec)-\xi_{\alpha,l}(\xvec)]{\partial\over\partial x_l}}
\int_V{d^3\xp\rho(\xpvec)(x_i-\xp_i)\over\vert\xvec-\xpvec\vert^3}
\nonumber\\
& &+\delta\int_V{d^3x F_{\alpha\beta,i}}.
\label{eq:v11}
\earay
Although we simplified the final answer by assuming that the fluids occupy identical
volumes and have proportional densities in the background state, we could not have
derived the correct perturbation of the first order virial theorem if we had not
allowed the volumes to differ.

For uniform ellipsoids, we can simplify the mutual gravitational term further.
First, we recognize that
\be
G\int_V{d^3\xp\rho(\xpvec)(x_i-\xp_i)\over\vert\xvec-\xpvec\vert^3}
={\partial\phi(\xvec)\over\partial x_i},
\ee
so we can write the gravitational term generally as
\be
f_\alpha f_\beta\int_V{d^3x\rho(\xvec)
[\xi_{\beta,l}(\xvec)-\xi_{\alpha,l}(\xvec)]{\partial^2\phi(\xvec)\over\partial x_l\partial x_i}}.
\ee
Second, recall that the potential at any interior point of a homogeneous ellipsoid
is (Theorem 3 in Chap. 3 of EFE)
\be
\phi(\xvec)=-\pi G\rho\biggl(I-\sum_{k=1}^3A_kx_k^2\biggr),
\ee
where $I$ is a constant; consequently
\be
{\partial^2\phi\over\partial x_l\partial x_i}=
2\pi G\rho A_i\delta_{il}.
\ee
Thus, the mutual gravitational contribution to the equation of motion for the
perturbed center of mass is
\be
2\pi G\rho^2A_if_\alpha f_\beta
\int_V{d^3x(\xi_{\beta,i}-\xi_{\alpha,i})}.
\ee
A sufficient condition for this to vanish is
\be
\int_V{d^3x\xivec_\alpha}=\int_V{d^3x\xivec_\beta},
\ee
which is just the statement that the perturbations have the same the center of mass,
specialized to the case of uniform density.

In Eq. (\ref{eq:v11}), we did not compute the variation in the last term. To do this,
let us write
\be
F_{SN,i}=-\rho_S\omega_S\beta_{ij}(u_{S,j}-u_{N,j});
\label{eq:fdefnew}
\ee
in the background state, the two fluids move together (and may even be stationary
in the rotating frame) so we have
\be
\delta\int_V{d^3x F_{\alpha\beta,i}}=-{\cal S}_{\alpha\beta}
{d\over dt}\biggl(f_S\int_V{d^3x\rho(\xvec)\omega_S\beta_{ij}
(\xi_{S,j}-\xi_{N,j})}\biggr),
\ee
where ${\cal S}_{\alpha\beta}=0$ if $\alpha=\beta$, 1 if $\alpha=S$ and
$\beta=N$, and $-1$ if $\alpha=N$ and $\beta=S$, and [see Eq. (\ref{eq:fdef})]
$
\beta_{ij}=\beta\delta_{ij}+\beta^\prime\epsilon_{ijm}\nu_m
+(\bdp-\beta)\nu_i\nu_j
$.
It is clear that the centers of mass of the two fluids remain stationary
if the fluid displacements are identical. However, there may be other
conditions under which they remain stationary.  For example, the
integrated mutual friction force will be zero as long as
\be
f_S\int_V{d^3x\rho(\xvec)\omega_S\beta_{ij}
(\xi_{S,j}-\xi_{N,j})}=0;
\ee
for a background with uniform density, vorticity and frictional
coupling coefficients, this is guaranteed if
\be
\int_V{d^3x\xivec_S}=\int_V{d^3x\xivec_N},
\label{eq:dispcond}
\ee
which is less restrictive than the requirement of identical displacements.
We found the same condition for the vanishing of the integrated, mutual
gravitational force for perturbations of uniformly dense ellipsoids.
Equation (\ref{eq:dispcond}) may be true, in fact, for all of the perturbations
considered in our paper, since both sides may vanish identically.

Next, consider variations of the second order virial equation. Most of
the terms are varied exactly as for single fluids; one exception is
\baray
\delta{\cal M}_{\alpha\beta,ij}&=&
-Gf_\alpha f_\beta\biggl\{\int_V{d^3x\rho(\xvec)\xi_\alpha(\xvec){\partial\over
\partial x_l}}\int_V{d^3\xp\rho(\xpvec)(x_i-\xp_i)(x_j-\xp_j)\over
\vert\xvec-\xpvec\vert^3}\nonumber\\
& &+\int_V{d^3x\rho(\xvec)[\xi_{\alpha,l}(\xvec)-\xi_{\beta,l}(\xvec)]
{\partial\over\partial x_l}}\int{d^3\xp\rho(\xpvec)(x_i-\xp_i)\xp_j\over
\vert\xvec-\xpvec\vert^3}\biggr\},
\earay
where we have specialized to backgrounds with proportional densities and
identical bounding volumes.
The first term in the brackets can be combined with $\delta{\cal M}_{\alpha,ij}$
and we find
\baray
\delta{\cal M}_{\alpha,ij}+(1-\delta_{\alpha\beta})\delta{\cal M}_{\alpha\beta,ij}=
-Gf_\alpha\int_V{d^3x\rho(\xvec)\xi_\alpha(\xvec){\partial\over
\partial x_l}}\int_V{d^3\xp\rho(\xpvec)(x_i-\xp_i)(x_j-\xp_j)\over
\vert\xvec-\xpvec\vert^3}\nonumber\\
-Gf_\alpha f_\beta\int_V{d^3x\rho(\xvec)[\xi_{\alpha,l}(\xvec)-\xi_{\beta,l}(\xvec)]
{\partial\over\partial x_l}}\int_V{d^3\xp\rho(\xpvec)(x_i-\xp_i)\xp_j\over
\vert\xvec-\xpvec\vert^3}.
\earay
The last equation can be written more compactly in terms of the functions
\be
{\cal B}_{ij}\equiv G\int_V{d^3\xp\rho(\xvec')(x_i-\xp_i)(x_j-\xp_j)\over
\vert\xvec-\xpvec\vert^3}\qquad
{\partial{\cal D}_j\over\partial x_i}\equiv
-G\int_V{d^3\xp\rho(\xpvec)\xp_j(x_i-\xp_i)\over\vert\xvec-\xpvec\vert^3}
\label{ddef}
\ee
defined in EFE, Chap. 2, Eqs. (14) and (27); using them, we find
\be
\delta{\cal M}_{\alpha,ij}+(1-\delta_{\alpha\beta})\delta{\cal M}_{\alpha\beta,ij}=
-f_\alpha\int_V{d^3x\rho\xi_{\alpha,l}{\partial{\cal B}_{ij}\over\partial x_l}}
+f_\alpha f_\beta\int_V{d^3x\rho(\xi_{\alpha,l}-\xi_{\beta,l})
{\partial^2{\cal D}_j\over\partial x_l\partial x_i}}.
\ee
We can rewrite this last result using EFE, Chap. 2 Eq. (28)
i.e.
\be
{\partial{\cal D}_j\over\partial x_i}={\cal B}_{ij}-x_j{\partial\phi\over\partial x_i};
\ee
with this substitution we get
\baray
\delta{\cal M}_{\alpha,ij}+(1-\delta_{\alpha\beta})\delta{\cal M}_{\alpha\beta,ij}&=&
-f_\alpha\int_V{d^3x\rho\xi_{\alpha,l}{\partial{\cal B}_{ij}\over\partial x_l}}
\nonumber\\
&+&f_\alpha f_\beta\int_V{d^3x\rho(\xi_{\alpha,l}-\xi_{\beta,l})
\biggl({\partial {\cal B}_{ij}\over\partial x_l}
-\delta_{lj}{\partial\phi\over\partial x_i}
-x_j{\partial^2\phi\over\partial x_l\partial x_i}\biggr)}.
\nonumber\\
\earay
For the uniform ellipsoids [EFE, Chap. 3, Eqs. (125) and (126)],
\be
{{\cal D}_j\over\pi G\rho}=a_j^2x_j\biggl(A_j-\sum_{k=1}^3A_{jk}x_k^2
\biggr)
\qquad
{{\cal B}_{ij}\over\pi G\rho}=2B_{ij}x_ix_j
+a_i^2\delta_{ij}\biggl(A_i-\sum_{i=1}^3A_{ik}x_k^2\biggr),
\ee
\baray
(\pi G\rho)^{-1}{\partial^2{\cal D}_j\over\partial x_l \partial x_i}&=&
-2a_j^2(\delta_{ij}x_lA_{jl}+\delta_{jl}x_iA_{ji}+\delta_{il}x_jA_{ji})
\nonumber\\
&=&-2a_j^2[A_{il}(\delta_{ij}x_l+\delta_{jl}x_i)+\delta_{il}x_jA_{ji}],
\nonumber\\
(\pi G\rho)^{-1}{\partial{\cal B}_{ij}\over\partial x_l}
&=&2B_{ij}(\delta_{il}x_j+\delta_{jl}x_i)
-2a_i^2\delta_{ij}A_{il}x_l.
\earay
Using these results and the definitions
[EFE, Chap. 2, Eqs. (122), (124), and (125)]
\be
V_{\alpha,i;j}\equiv\int_V{d^3x\rho\xi_{\alpha,i}x_j}\qquad
V_{\alpha,ij}=V_{\alpha,i;j}+V_{\alpha,j;i},
\label{eq:vdef}
\ee
we find
\baray
{\delta{\cal M}_{\alpha,ij}+(1-\delta_{\alpha\beta})\delta{\cal M}_{\alpha\beta,ij}\over
\pi G\rho}=
-f_\alpha\biggl(2B_{ij}V_{\alpha,ij}-a_i^2\delta_{ij}\sum_{l=1}^3A_{il}V_{\alpha,ll}
\biggr)\nonumber\\
-a_j^2f_\alpha f_\beta\biggl[2A_{ij}(V_{\alpha,ij}-V_{\beta,ij})
+\delta_{ij}\sum_{l=1}^3A_{il}(V_{\alpha,ll}-V_{\beta,ll})\biggr].
\label{eq:mutgrav}
\earay
It is possible to write this more compactly, but the form of Eq. (\ref{eq:mutgrav})
makes clear which terms depend on the differences between the displacements of the
two fluids and which do not.

The other new moment we need is
\be
\delta{\cal F}_{\alpha\beta,ij}=\delta\int_{V_\alpha}{d^3xx_jF_{\alpha\beta,i}}.
\ee
Since the two fluids move together in the unperturbed state, this moment is
first order in the perturbations at the largest. We then find
\be
\delta\int_{V_\alpha}{d^3xx_jF_{\alpha\beta,i}}=
-{\cal S}_{\alpha\beta}f_S\int_V{d^3x\rho\omega_Sx_j\beta_{ik}\biggl(
{d\xi_{S,k}\over dt}-{d\xi_{N,k}\over dt}\biggr)}.
\label{mfdef}
\ee
For perturbations of uniform ellipsoids, $\omega_S$ and $\rho$ are independent
of position in the unperturbed background, and we may also assume that $\beta_{ij}$
is constant; then,
\be
\delta\int_{V_\alpha}{d^3xx_jF_{\alpha\beta,i}}
=-{\cal S}_{\alpha\beta}f_S\omega_S
\beta_{ik}\biggl[{d\over dt}\biggl(\int_V{d^3x\rho
x_j(\xi_{S,k}-\xi_{N,k})}\biggr)
-\int_V d^3x\rho u_j(\xi_{S,k}-\xi_{N,k})\biggr].
\ee
For backgrounds in which there are no fluid motions, the last term is absent and
\be
\delta\int_{V_\alpha}{d^3xx_jF_{\alpha\beta,i}}=-{\cal S}_{\alpha\beta}f_S\rho\omega_S
\beta_{ik}\biggl({dV_{\alpha,k;j}\over dt}-{dV_{\beta,k;j}\over dt}\biggr),
\ee
using the definition in Eq. (\ref{eq:vdef}).

When there are no fluid motions of the unperturbed star in the rotating frame,
the second order virial equations are
\baray
f_S{d^2V_{S,i;j}\over dt^2}&=&2\epsilon_{ilm}\Omega_mf_S{dV_{S,l;j}\over dt}
+\Omega^2f_SV_{S,ij}-\Omega_i\Omega_kf_SV_{S,kj}+\delta_{ij}\delta\Pi_S
\nonumber\\
&-&f_S\pi G\rho\biggl(2B_{ij}V_{S,ij}-a_i^2\delta_{ij}\sum_{l=1}^3A_{il}V_{S,ll}
\biggr)
\nonumber\\
&-&a_j^2f_S f_N\pi G\rho\biggl[2A_{ij}(V_{S,ij}-V_{N,ij})
+\delta_{ij}\sum_{l=1}^3A_{il}(V_{S,ll}-V_{N,ll})\biggr]
\nonumber\\
&-&f_S\omega_S
\beta_{ik}\biggl({dV_{S,k;j}\over dt}-{dV_{N,k;j}\over dt}\biggr),
\nonumber\\
f_N{d^2V_{N,i;j}\over dt^2}&=&2\epsilon_{ilm}\Omega_mf_N{dV_{N,l;j}\over dt}
+\Omega^2f_NV_{N,ij}-\Omega_i\Omega_kf_NV_{N,kj}+\delta_{ij}\delta\Pi_N
\nonumber\\
&-&f_N\pi G\rho\biggl(2B_{ij}V_{N,ij}-a_i^2\delta_{ij}\sum_{l=1}^3A_{il}V_{N,ll}
\biggr)
\nonumber\\
&-&a_j^2f_N f_S\pi G\rho\biggl[2A_{ij}(V_{N,ij}-V_{Sij})
+\delta_{ij}\sum_{l=1}^3A_{il}(V_{N,ll}-V_{S,ll})\biggr]
\nonumber\\
&-&f_S\omega_S
\beta_{ik}\biggl({dV_{N,k;j}\over dt}-{dV_{S,k;j}\over dt}\biggr).
\earay
We can replace these equations with a different set by defining
\be
V_{i;j}\equiv f_SV_{S,i;j}+f_NV_{N,i;j},\qquad
U_{i;j}\equiv V_{S,i;j}-V_{N,i;j}.
\ee
In terms of these new quantities we find Eqs. (\ref{meanflow_basic})
and (\ref{eq:differentmom1}).


\begin{center}
\includegraphics[height=6.2in,width=6.2in]{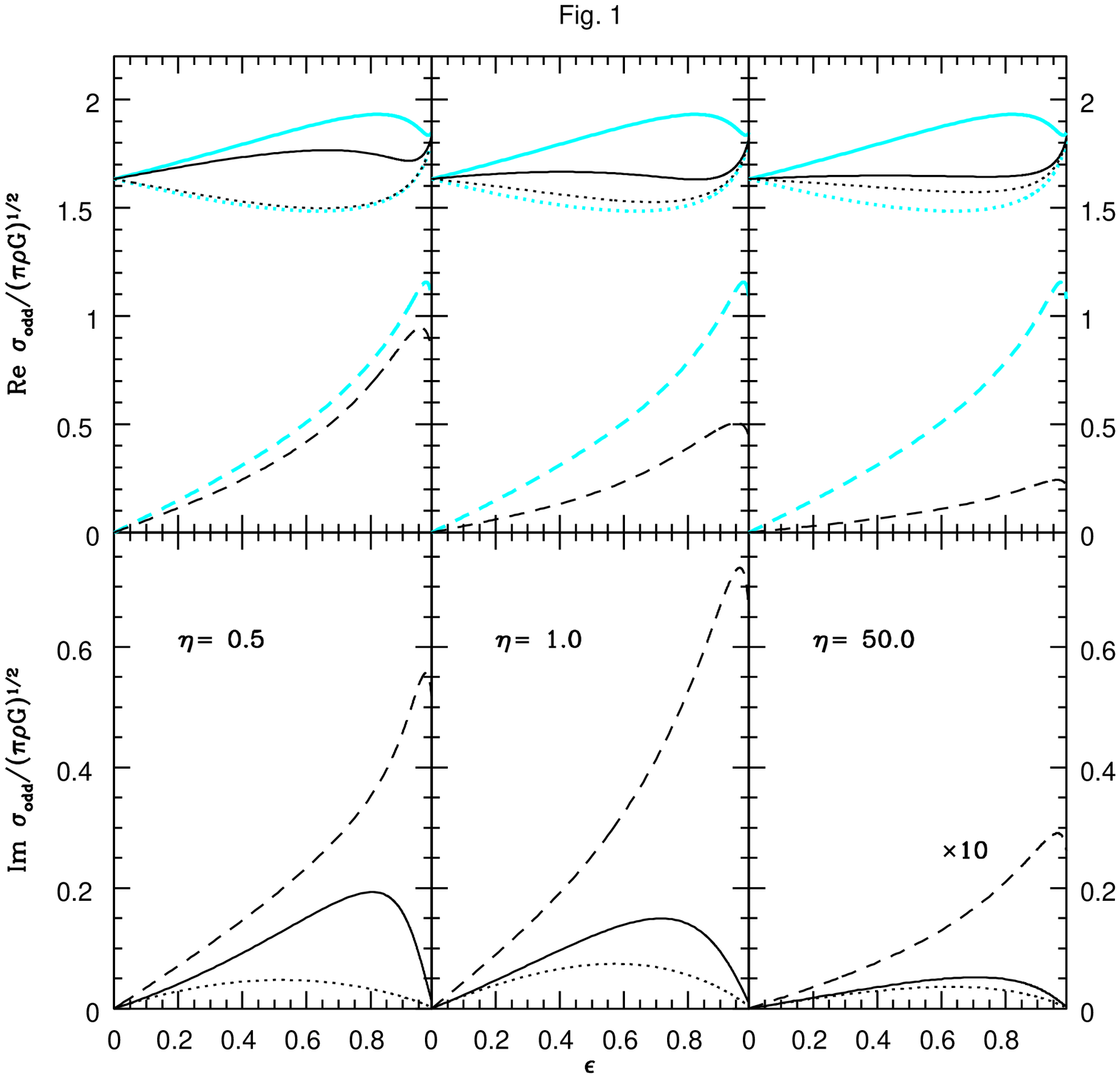}
\end{center}
{\footnotesize{Fig 1: The real (upper panel) and imaginary (lower panel)
parts of the relative transverse-shear modes of a superfluid Maclaurin
spheroid as a function of eccentricity for three values of $\eta = 0.5,
~1, ~50$. The $\eta$ parameter is scaled in units of $\omega_S\rho_S$.
The imaginary parts of the modes for $\eta = 50$ are magnified
by a factor of 10. The grey lines show the frictionless solutions. To relate
the $\beta$-coefficients to the rescaled $\tilde\beta$-coefficients
we have set $f_S/f_N = 0.2$. The results are insensitive to the choice
of this ratio.
}}
\label{fig1}

\begin{center}
\includegraphics[height=6.2in,width=6.2in]{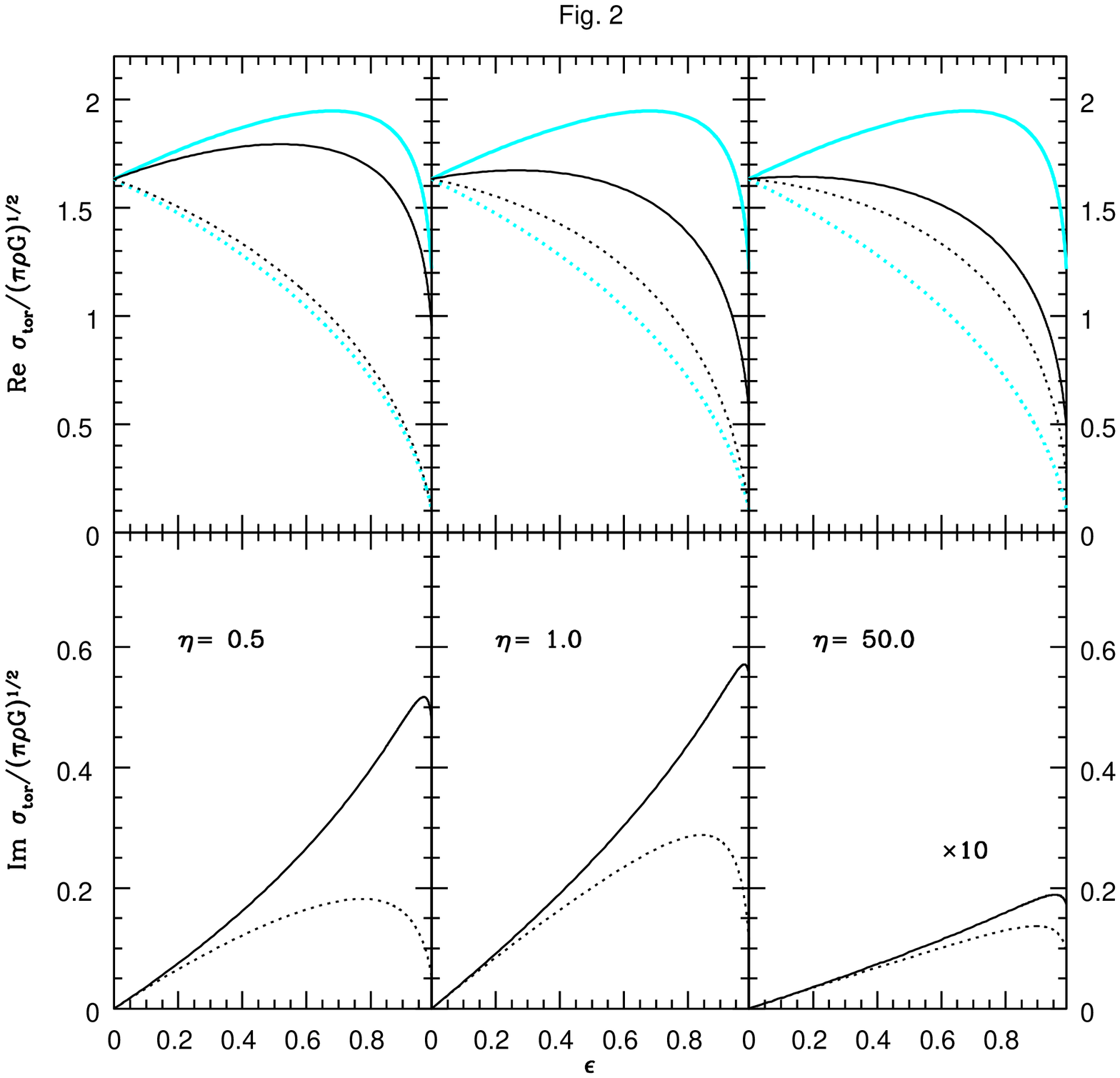}
\end{center}
{\footnotesize{Fig 2: The relative  toroidal modes of  superfluid Maclaurin
spheroids.  Conventions  are the same as in Fig. 1.
}}
\label{fig2}

\begin{center}
\includegraphics[height=6.2in,width=6.2in]{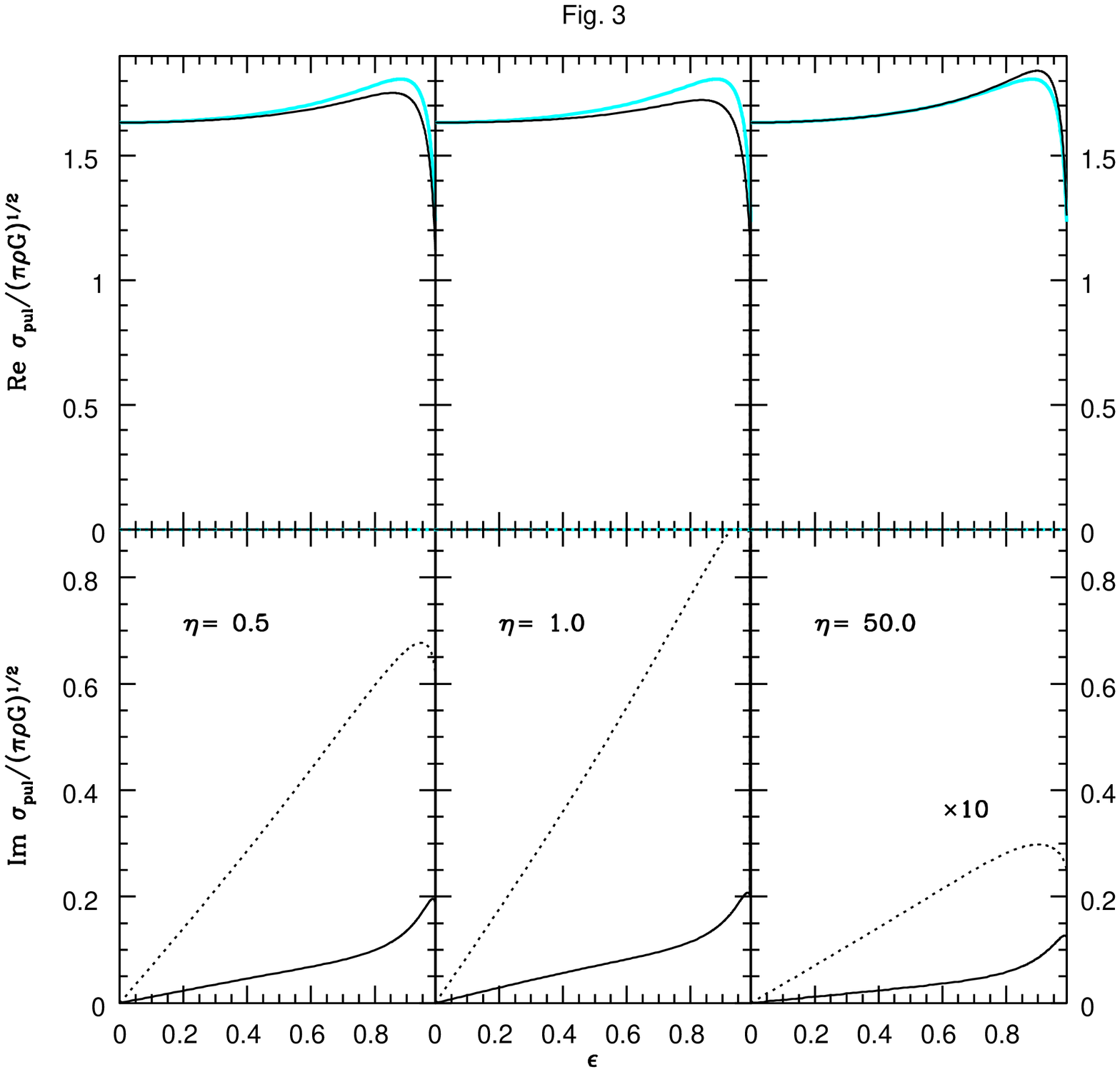}
\end{center}
{\footnotesize{Fig 3: The relative pulsation modes of  superfluid Maclaurin
spheroids.  Conventions  are the same as in Fig. 1.
}}
\label{fig3}

\begin{center}
\includegraphics[height=6.2in,width=6.2in]{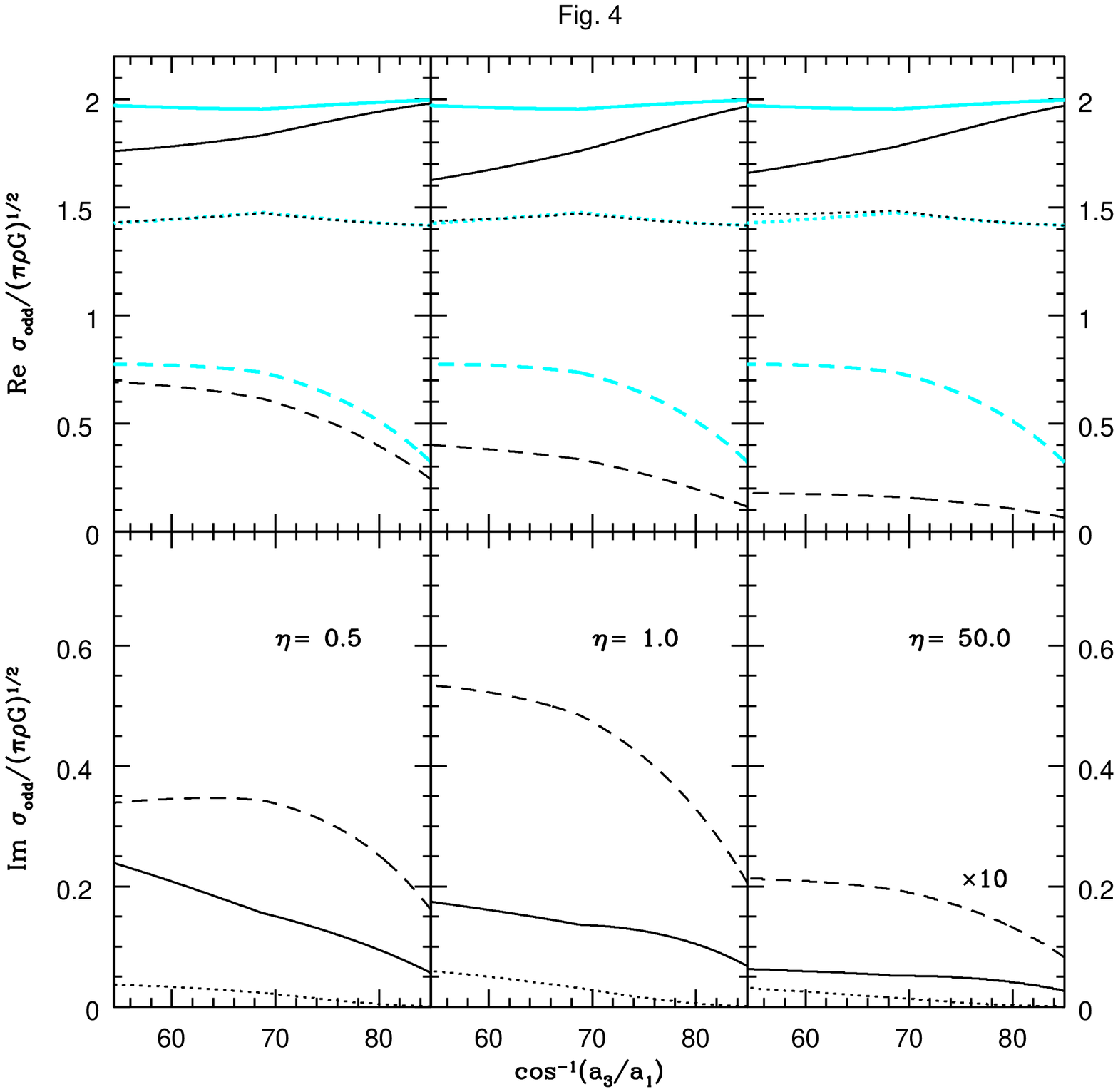}
\end{center}
{\footnotesize{Fig 4: The relative  odd-parity modes of superfluid Jacobi
ellipsoids as a function of ${\rm cos}^{-1}(a_3/a_1)$.
Conventions  are the same as in Fig. 1.
}}
\label{fig4}

\begin{center}
\includegraphics[height=6.2in,width=6.2in]{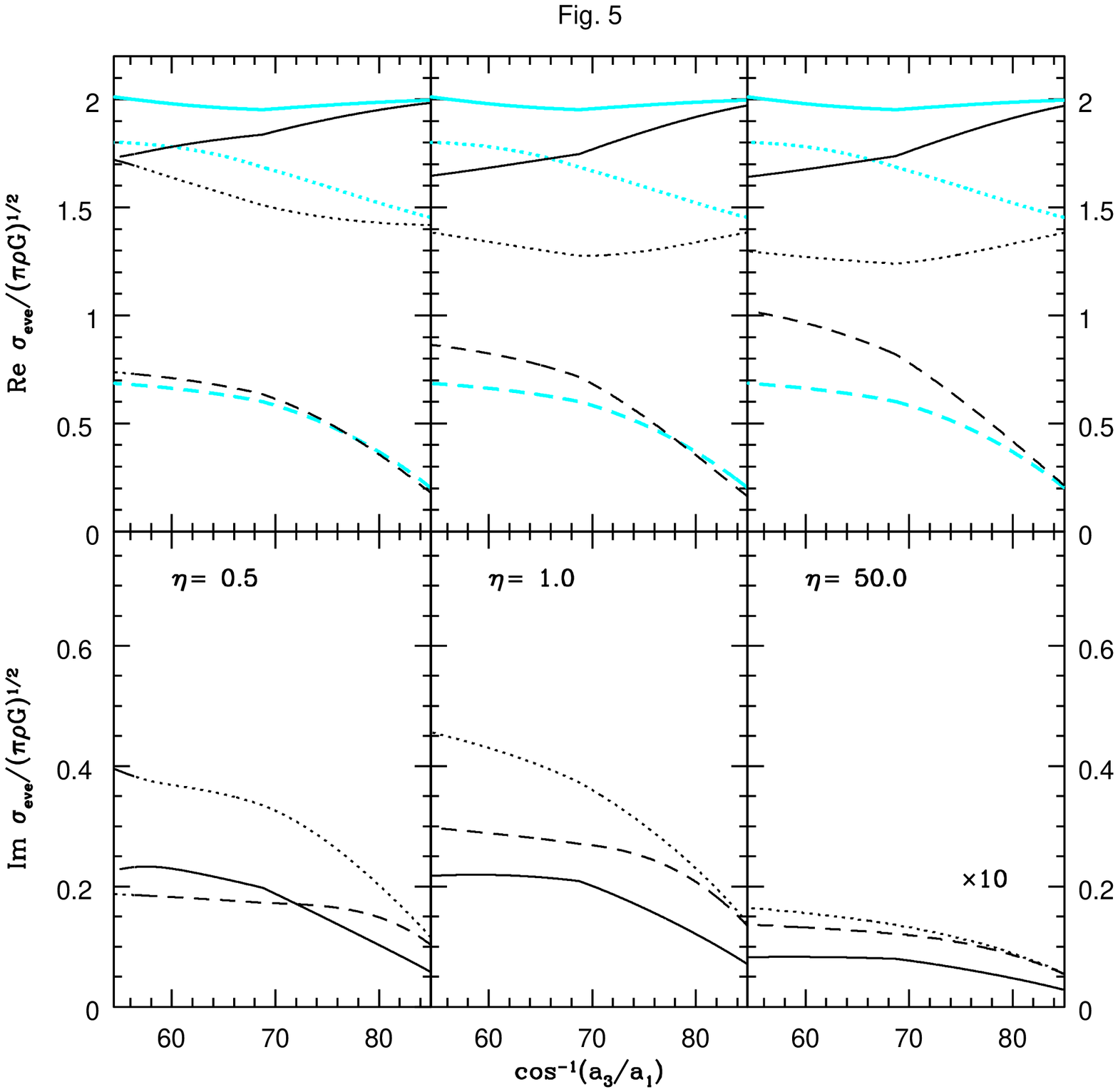}
\end{center}
{\footnotesize{Fig 5: The relative  even-parity modes of  superfluid Jacobi
ellipsoids as a function of ${\rm cos}^{-1}(a_3/a_1)$.
 Conventions  are the same as in Fig. 1.
}}
\label{fig5}

\begin{center}
\includegraphics[height=6.2in,width=6.2in]{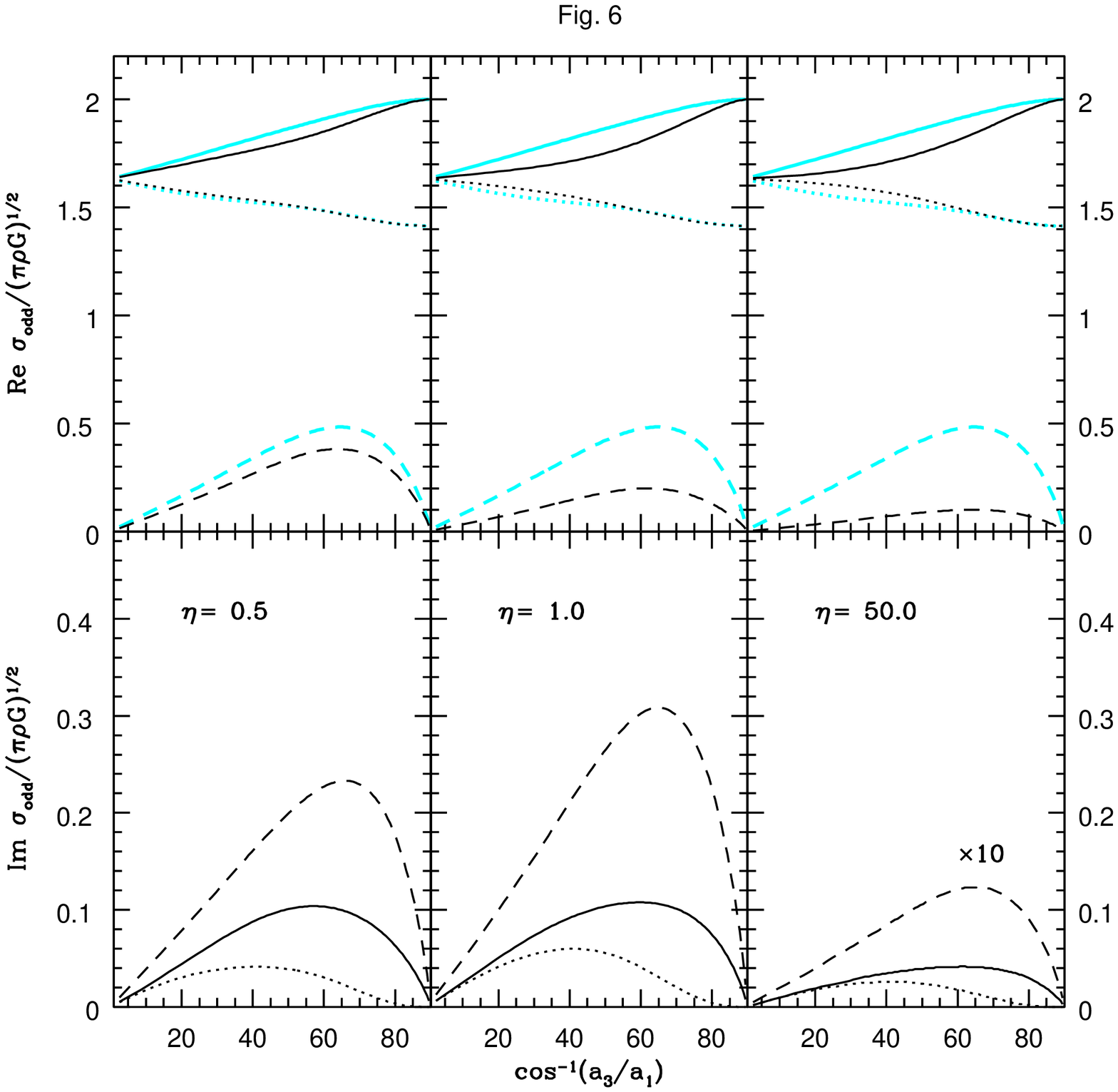}
\end{center}
{\footnotesize{Fig 6: The relative  odd-parity modes of  superfluid Roche
ellipsoids as a function of ${\rm cos}^{-1}(a_3/a_1)$ for $P=1$.
  Conventions  are the same as in Fig. 1.
}}
\label{fig6}

\begin{center}
\includegraphics[height=6.2in,width=6.2in]{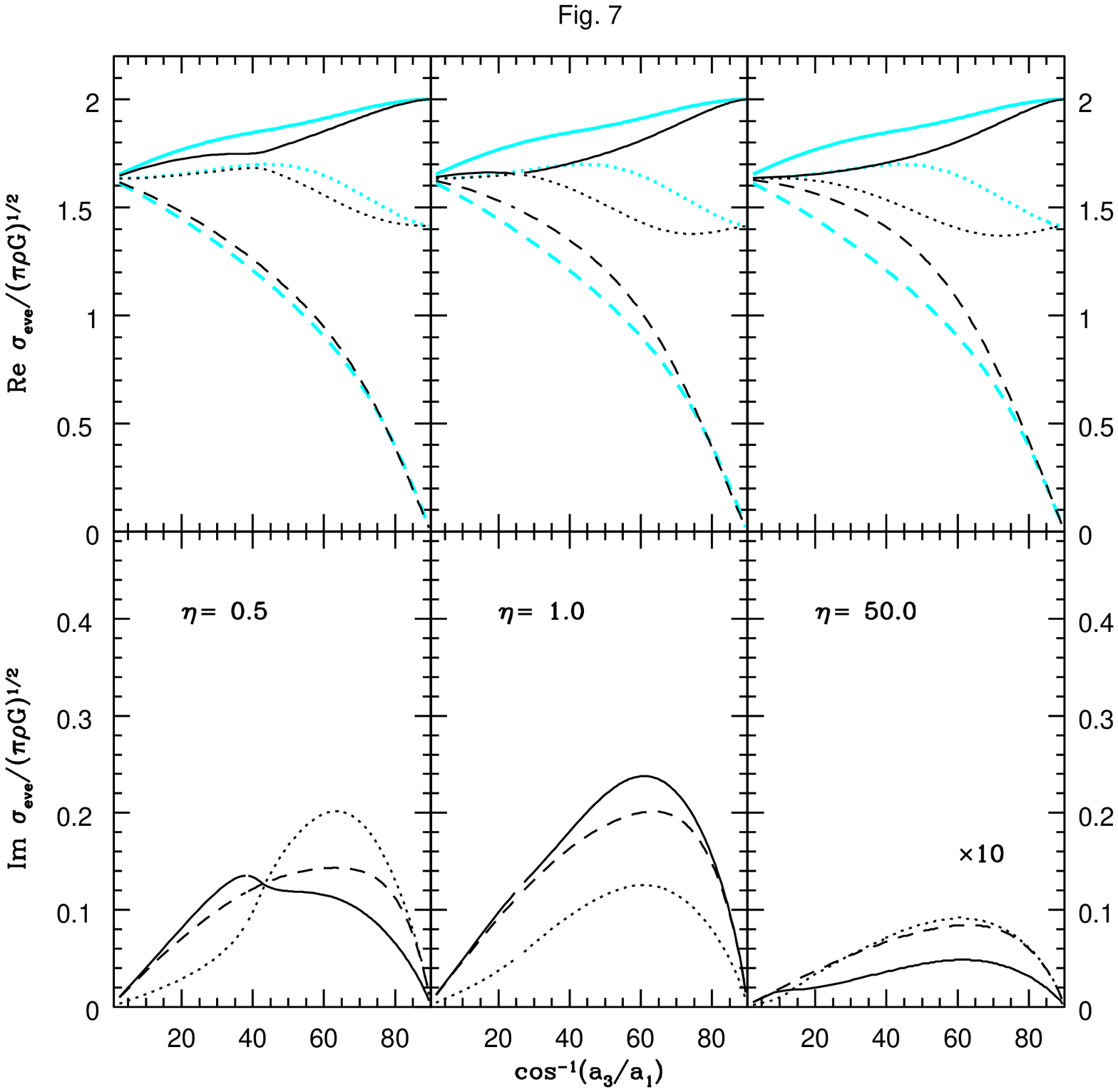}
\end{center}
{\footnotesize{Fig 7: The relative  even-parity modes of superfluid Roche
ellipsoids as a function of ${\rm cos}^{-1}(a_3/a_1)$ for $P=1$.
 Conventions  are the same as in Fig. 1.
}}
\label{fig7}
\end{document}